\documentclass[prb,reprint]{revtex4-2}
\usepackage{graphicx}
\usepackage{dcolumn}
\usepackage{bm}
\usepackage[svgnames]{xcolor}
\usepackage{mathtools}
\usepackage{dsfont}
\usepackage{url}
\usepackage{hyperref}
\usepackage[version=4]{mhchem}

\usepackage{ulem}
\usepackage{soul}

\begin{document}
	
	\preprint{APS/123-QED}
	
	\title{Disorder induced topological phase transition in a driven Majorana chain}
	
	\author{Henry Ling}
	\author{Philip Richard$^1$}
	\author{Saeed Rahmanian Koshkaki$^2$}
	\author{Michael Kolodrubetz$^2$}
	\author{Dganit Meidan$^{3,4}$}
	\author{Aditi Mitra$^5$}
	\author{T. Pereg-Barnea$^1$}
	
	\affiliation{$^1$Department of Physics, McGill University, Montr\'eal, Qu\'ebec, Canada H3A 2T8}
	\affiliation{$^2$Department of Physics, The University of Texas at Dallas, Richardson, Texas 75080, USA}
	\affiliation{$^3$Department of Physics, Ben-Gurion University of the Negev, Beer-Sheva 84105, Israel}
     \affiliation{$^4$Université Paris-Saclay, CNRS, Laboratoire de Physique des Solides, 91405, Orsay, France}
	\affiliation{$^5$Center for Quantum Phenomena, Department of Physics, New York University, 726 Broadway, New York, NY, 10003, USA}

	\date{\today}
	
	\begin{abstract}
		We study a periodically driven one dimensional Kitaev model in the presence of disorder.  In the clean limit our model exhibits four topological phases corresponding to the existence or non-existence of edge modes at zero and $\pi$ quasienergy.  When potential disorder is added, the system parameters get renormalized and the system may exhibit a topological phase transition.  When starting from the Majorana $\pi$ Mode (MPM) phase, which hosts only edge Majoranas with 
		quasienergy $\pi$, disorder induces a transition into a neighboring phase with both $\pi$ and zero modes on the edges.  We characterize the disordered system using (i) exact diagonalization (ii) Arnoldi mapping onto an effective tight binding chain and (iii) topological entanglement entropy.
	\end{abstract}
	
	\maketitle

	\section{Introduction}
	The ability to drive a system in and out of various topological phases has attracted a lot of attention in recent years.  While topological properties may be altered by chemical doping and structural manipulation, the idea of controlling topology using an external knob is extremely appealing.  Using a time periodic perturbation such as light is therefore highly desirable and has been studied theoretically by many authors \cite{Gu,Lindner1,Dehghani1,Dehghani2,Farrell1,Farrell2,Kundu}.  Moreover, several experiments point to the feasibility of this kind of Floquet band engineering, demonstrating the creation of side bands, band renormalization \cite{Wang_2013, Lee22} and even light-induced Berry curvatures \cite{McIver_2020}.  Analogues of Floquet topological systems can also be found in photonic crystals \cite{Rechtsman_2013,Mukherjee_2021,Guglielmon_2018}.  
	
	Among a variety of topological systems, of special interest is the one dimensional topological superconductor.  It can host unpaired Majorana fermions at its edges and is therefore sought after as a potential building block of qubits, taking advantage of the non-Abelian nature of edge Majorana modes \cite{Alicea_2011,Plugge_2010}.  The existence of unpaired Majoranas was predicted in the Kitaev model of spinless fermions \cite{Kitaev_2001} and possible realizations consisting of wires with spin orbit coupling were proposed subsequently \cite{Lutchyn_2010,Oreg_2010}. The Kitaev chain  is simple and versatile as it can be written in the language of superconductivity, in terms of Majorana operators or as a spin chain.  While these descriptions are analogous they may inspire different realizations 
 
 The topology of the Kitaev chain is not only apparent from the existence of edge modes, but it can also be characterized by a topological $Z_2$ invariant \cite{Ryu10}.  Similarly, one can use the entanglement entropy (EE) as an indicator of topology.  This has been shown in intrinsic topological systems, 
	which are  characterized by a universal constant, added to the usual area law \cite{Preskill06,Levin2006}, and in symmetry protected states, where topological contributions to the EE appear in the area law itself or in subleading terms \cite{Borchmann_2014,Borchmann_2016,Borchmann_2017}.  Specifically, in the Kitaev chain, a topological entanglement entropy can be found through differences of the entanglement entropy of different partitions \cite{Zeng2015,Zeng_2016,Fromholz_2020}.
	
	Given the interest in topological superconductors and the demand for adjustable knobs, the study of periodically driven Majorana or spin chains is a natural choice.  Most strikingly, the phase diagram of the driven Kitaev chain is richer than that of the equilibrium model \cite{Khemani_2016,Jiang_2011}.  At equilibrium the Kitaev chain contains two parameters - the pairing and the chemical potential, whose strengths are measured with respect to the hopping amplitude. The Kitaev chain has two phases - it is either trivial, without any unpaired Majoranas, or topological with exponentially decoupled and localized Majorana fermions at its edges.  When driving the system, the frequency of the drive adds another parameter that can control the topology of the system.  Consequently, the driven system may host not only zero Majorana modes (MZMs) at its edges but also $\pi$ Majorana modes (MPMs).  
	
	In a floquet driven system, the reduction of symmetry to discrete time translation forces the energy to be conserved only modulo the drive frequency, $\Omega$. The resulting quasienergy  $\epsilon$ resides within a Floquet zone, such that  $-\Omega/2<\epsilon\leq\Omega/2$ where $\hbar=1$ throughout. 
	In terms of the period $T=2\pi/\Omega$ the zone is defined by $\epsilon T \in(-\pi,\pi]$.   MPMs are at the edges of the Floquet zone and, just like MZMs, reside in a particle hole symmetric quasienergy.  MPMs are hence Majorana modes which acquire a phase shift $\pi$  at every drive cycle.  Like MZMs, edge MPMs are also unpaired and can be used for braiding purposes.  They can coexist with MZMs as the different quasienergies ensure orthogonality, as long as the time-periodicity is (at least approximately) preserved.  Therefore, the system has now four phases corresponding to the existence of MZMs and MPMs.  Moreover, the MZMs are orthogonal to the MPMs as long as the system is periodic, at least approximately.  This leads to further advantages like the ability to exchange Majoranas on a single wire \cite{Bauer_2019}.  It is also interesting to note that allowing for longer range couplings can lead to a richer phase diagram with multiple zero modes and/or $\pi$ modes 
	\cite{Obuse13,Delplace14,Yates17,Yates18}.
	
	Disorder, inevitable in any physical system, is another issue that needs to be addressed.  Majorana modes are topologically protected against weak disorder, since they are exponentially localized as a function of the system size. However, disorder is known to make MZMs less localized at the system’s boundary, and can drive a topological phase transition at a critical disorder strength \cite{Motrunich2001,Brouwer2011,Brouwer2011a,Akhmerov2011,Rieder2012,Rieder2013,DeGottardi2013, Pan2020zerobiasconductance, Pan2021DisorderMZM, das2023search}. 
	In addition, disorder may lead to localization of bulk modes which otherwise could couple to edge modes.  This may decrease the coupling between edge modes and cause the system to be more robust. In fact, some authors claim that many body localization can protect Majorana fermions from hybridizing, increasing the fidelity of Majorana based qubits \cite{Huse_2013,Laflorencie_2020}.  It should be noted, however, that this claim is challenging to verify numerically \cite{Kells2018}.
	
	In driven systems, many recent works showed the robustness of Floquet topological phases and transport against disorder~\cite{shapiro2019strongly, Shtanko2018StabilityTopoDisorder, Farrell2018EdgeStateDisorder, Titum2015DisorderInduceTopology, PhysRevLett.124.190601}. A less investigated scenario is the stability of the MPM when the system has strong disorder.  In this work, we further study the effect of disorder on Floquet topological phases, specifically the MPM phase, by studying the periodically driven Kitaev chain. Surprisingly, we find that increasing the disorder strength drives a topological phase transition: The disorder  
 does not significantly affect the MPMs while enabling the emergence of MZMs.
    
    The model is presented in section \ref{sec:Model}. We first analyze the model with the Arnoldi method \cite{Arnoldi} in Sec.~\ref{sec:Arnoldi}, complimenting it by exactly diagonalizing the Floquet unitary, and showing the localization of eigenstates throughout the spectrum.  Besides the existence of Majorana edge modes at zero and/or $\pi$ quasienergy, we also characterize the topology of the system calculating the topological part of the entanglement entropy in Sec.~\ref{sec:TEE}. Exact solutions for the edge modes in the presence of disorder are presented in Sec.~\ref{sec:exact}, and are used to identify the topological phase transition, and complementing the analysis from the entanglement entropy. We conclude in Sec.~\ref{sec:Conclusions}.
	
	\section{Model of the disordered and driven Kitaev chain}\label{sec:Model}
	We choose a stroboscopic drive such that the period $T$ is divided between two Hamiltonians.  We write our Hamiltonian in the fermionic language and remind the reader that it is equivalent to a spin chain (up to boundary terms that do not matter for open boundary conditions and/or for a given parity sector).  The Floquet time evolution operator of a single cycle can be written as:
	\begin{eqnarray}\label{eq:Unitary}
		U(T)=\exp\left(-i {{\cal H}_2} {T\over 2}\right)\exp\left(-i {{\cal H}_1} {T\over 2}\right),
	\end{eqnarray}
	where $T$ is the drive period and we 
 use units where $\hbar=1$. Note that this choice of driving is convenient for numerical simulations as ${\cal H}_1$ and ${\cal H}_2$ are time independent Hamiltonians, and contain different terms of the Kitaev Hamiltonian of spinless fermions with p-wave pairing.  On a real space lattice this gives:
	\begin{eqnarray}\label{eq:Hamiltonians}
		{\cal H}_1 &=& \sum_{j=1}^{L-1}\left(-w c^\dagger_j c_{j+1} + \Delta c_jc_{j+1}+h.c\right),\nonumber \\
		{\cal H}_2 &=& -\sum_{j=1}^N\mu_j c^\dagger_j c_j, 
	\end{eqnarray}
where $j$ is the site index, $L$ is the total number of sites on the chain, $w$ is the hopping amplitude, $\Delta$ is the uniform pairing amplitude, and the disorder is only added to the chemical potential as $\mu_j = \Bar{\mu}+\delta \mu_i$. 
		
	The disorder can be expressed in the form $\delta \mu_i = \sigma u_i$, where $u_i$ is the disorder profile and $\sigma$ is the disorder strength.  We produce disorder profiles, $\{u_i\}$, by choosing a series of $L$ random numbers, under the constraints of a zero mean and unit variance:
	$\sum_i  \mu_i/L =\bar{\mu}$, $\sum_i (\mu_i-\bar{\mu})^2/L= \sigma^2 $. In the large chain limit, the chemical potential therefore follows a Gaussian distribution with $ \bar{\mu}$ mean and $\sigma^2 $ variance. 
	
	Note that this model is equivalent, through the Jordan Wigner transformation, to a driven spin chain with:
	\begin{eqnarray}\label{eq:SpinChain}
		{\cal H}_1 &=& \sum_j\left(J_x \sigma^x_j \sigma^x_{j+1}+ J_y \sigma^y_j \sigma^y_{j+1}\right),\nonumber \\
		{\cal H}_2 &=& \sum_jg_j\sigma_j^z,  
	\end{eqnarray}
	where $\sigma_i^\alpha$ are Pauli matrices, $J_x = (\Delta+w)/2$, $J_y=(\Delta-w)/2$ and $g=\mu/2$.

	The above model was studied in the clean limit by various authors, with and without integrability breaking perturbations \cite{Khemani_2016,Yates_2019,Bauer_2019,Yates_PRL_2020,Yates_2020,Yates_2021,Yates_2022,Yeh_2023}, while the undriven case has also been studied in the presence of disorder \cite{Motrunich2001,Brouwer2011,Akhmerov2011,Rieder2012,Rieder2013,DeGottardi2013,Pientka_2013,Gergs_2016,Hegde_2016,Pan_2021}.   In this work we study the driven and disordered system.

	\section{Krylov space mapping using the Arnoldi method and exact diagonalization}\label{sec:Arnoldi}
	We characterize the edge mode of our system by studying the time evolution of a Majorana operator, initially localized to the left edge of the chain.  This `seed operator' is defined as $\gamma_1 = c_1+c_1^\dagger$ 
	where $1$ is the left most site of our chain. We apply  the Floquet unitary $U(T)$ defined in Eq.~\eqref{eq:Unitary} $n$ times and calculate the resulting operator's overlap with the initial operator, $\gamma_1$.  This results in the definition of the ``infinite temperature'' autocorrelation function $A_{\infty}$:
	\begin{eqnarray}
		A_{\infty}(nT) = {1\over 2^L}{\rm Tr}[\gamma_1(nT)\gamma_1(0)],
	\end{eqnarray}
	where $\gamma_1(0)=\gamma_1$  and 
	\begin{eqnarray}
		\gamma(nT) = U^\dagger(nT)\gamma_1U(nT)=(U^\dagger)^n\gamma_1U^n(T).
	\end{eqnarray}
	
$A_\infty$ being the autocorrelation function of the edge operator $\gamma_1$, captures how fast this operator spreads into the bulk, or equivalently, how information propagates into the bulk. In addition, being a trace over the entire Hilbert space, this quantity also averages out small fluctuations. In the absence of localized edge modes, $A_{\infty}$ will decay to zero rapidly in time. Conversely, in the presence of zero or $\pi $ modes,  and in the thermodynamic limit $L\rightarrow \infty$, after the initial transient, this quantity will reach a plateau value for infinite time because of the exact commutation/anti-commutation relation of the strong edge mode. For a finite size system however, $A_{\infty}$ after reaching the plateau will eventually decay to zero on a time scale that is exponential in the system size.
 
 One may analyze the zero and $\pi$ modes by studying the Heisenberg time-evolution of the operator which has an overlap with these modes, in Krylov subspace.  The advantage of the latter is that it maps the system onto a chain of non-interacting particles without pairing \cite{Recbook, Arnoldi}. For a non-interacting unitary, the Krylov mapping just maps one non-interacting problem to another non-interacting problem.  Nevertheless, it is a conceptually alternative way to study edge modes, and paves the way for clarifying the effects of disorder and/or interactions on these modes.  
	
	Following Yates {\it et al.} \cite{Yates_2021} we would like to map our problem to a tight binding chain in Krylov subspace.  We define a space of states $|O)$ in the Krylov subspace, which correspond to  operators in the Kitaev chain.  The scalar product in this subspace is defined by the trace:
	\begin{equation}
		(A|B) = \frac{1}{2^L} \text{Tr} \left[ A^\dagger B \right].
	\end{equation}
	The seed operator $\gamma_1$ is the first Krylov state: \(|1) = \gamma_1 \).  Subsequent states are obtained by successive applications of the Floquet time evolution operator, with the tight binding Krylov model encoded in the matrix $W$ defined as: 
	\begin{align}
		W |O) &= U^\dagger \hat{O} U,\\
		W^n|O) &= \left[U^\dagger\right]^n \hat{O}\left[U\right]^n.
	\end{align}
	Above, we have dropped the argument $T$ from the Floquet time evolution operator.
	The unitarity of $W$ follows from the unitarity of $U$:
	\begin{align}
		W^\dagger W |O) &= U (U^\dagger \hat{O} U) U^\dagger = \hat{O}.
	\end{align}
	We now outline the Arnoldi method \cite{Arnoldi}.
	Let \(|1) = \gamma_1 \). We find subsequent states by time evolving the state using $W$.  The resulting operator is a state in the Krylov subspace.  The part of this state which is orthogonal to all previous states is a new basis vector up to normalization.  In other words, assuming we have found the orthonormal
	basis vectors \(|n),|n-1),\dots \),  the next state \(|n+1)\) is found by:
	\begin{align}
		|n+1') &= W|n) - \sum_{l= 1}^{n}|l)(l|W|n)\nonumber\\
		&= W|n) - \sum_{l = 1}^n w_{l,n} |l)\label{eq:W1}\\
		&= \left[ 1 - \sum_{l = 1}^{n}|l)(l| \right] W |n) = P_nW |n).
	\end{align}
	Above $w_{l,n}= (l|W|n)$ and $P_n=1 - \sum_{l = 1}^{n}|l)(l|$ projects out overlaps with the previously
	calculated basis vectors.
	Following this we  normalize \(|n+1')\),
	\begin{equation}
		|n+1) = \frac{|n+1')}{\sqrt{(n+1'|n+1')}}.\label{eq:W2}
	\end{equation}
	We note that
	\begin{align}
		(n+1'|n+1') &= (n+1'|P_n W|n) = (n+1'|W |n),
	\end{align}
	since $P_n^2=P_n$ is a projector. Using Eq.~\eqref{eq:W2} the above becomes
	\begin{align}
		(n+1'|n+1') = \sqrt{(n+1'|n+1')}(n+1|W|n),
	\end{align}
	implying that
	\begin{align}
		\sqrt{(n+1'|n+1')} &= (n+1|W|n) = w_{n+1,n}.
		\label{eq:subd}
	\end{align}
		
	Rearranging Eq.~\eqref{eq:W1}, we arrive at,
	\begin{equation}
		W |n) = w_{n+1,n} |n+1) + \sum_{l = 1}^n w_{l,n} |l).
	\end{equation}
	Thus as we iterate through this Arnoldi algorithm, we find that
	$W$ has the upper Hessenberg form, i.e., a square matrix whose elements $w_{i,j}=0$ for $i>j+1$\cite{Horn_2013}. In addition,
	when the seed operator is Hermitian, all the elements of $W$ are real because under time-evolution
	a Hermitian operator stays Hermitian, with the matrix elements of $W$ simply denoting the weight
	of different Hermitian operators at a particular step in the iterative procedure. 
	
	The autocorrelation function in terms of $W$ therefore has the form
	\begin{equation}\label{ainfWN}
		A_\infty(nT) = (1|W^n|1).
	\end{equation}
	We note that in the case of a large space we might decide to truncate the derivation of $W$ before we exhaust the Krylov space (for the non-interacting Kitaev chain this would happen at $2L$ iterations).  Therefore $W$ will not be exactly unitary. However, if the truncation size $N$ is sufficiently large, the truncated $W$ reproduces the dynamics very well.
	The success of the approximation is a good indication
	that if the edge modes are sufficiently localized at the edge of the Krylov chain, the truncation scheme
	does not affect the physics. After obtaining the effective Arnoldi unitary, $W$, we can analyze the system by studying the spectrum and eigenmodes of $W$, which can be used towards computing the autocorrelation $A_\infty$.
	
\begin{figure}
		\centering
		\includegraphics[width=1.0\columnwidth]{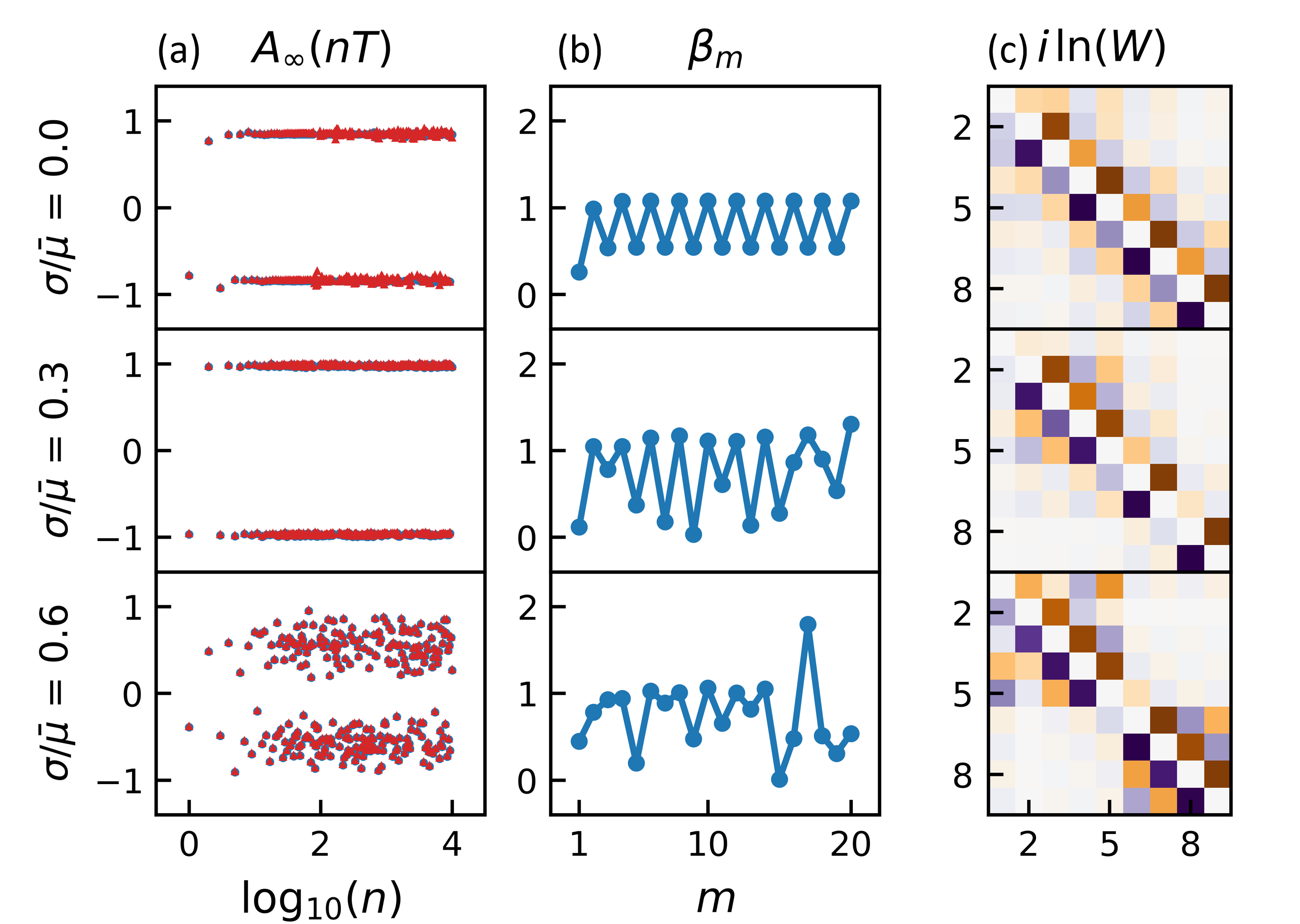}
		\caption{(a) A typical autocorrelation function $A_\infty(nT)$ as a function of $\log_{10}(n)$. The autocorrelation is calculated using both the matrix $W$ (red) as well as through exact diagonalization (blue).  The two methods agree and therefore the blue symbols are not visible. 
			The results are presented for a single disorder realization scaled such that the standard deviation is $0, 0.3, 0.6$ (from top to bottom). (b) The nearest neighbor hopping, $\beta_m$, as a function of the Krylov site index $m$, in the effective Krylov-space chain, given by the first diagonal next to the main one of the matrix $i\ln(W)$.  (c) The first few elements of $i\ln(W)$ represented by color.  Darker colors corresponds to larger elements.  In the above calculation we have used the model in Eq.~\eqref{eq:Unitary} with parameters $w=\Delta=1$ and $\bar\mu=0.6$.  $\sigma$ is the standard deviation of the disorder, measured relative to the average chemical potential $\bar \mu$.  
		}
		\label{fig:Ainf}
	\end{figure}

	We begin our study with a system which falls in the $\pi$-mode phase in the clean limit.  This means that without any disorder we expect the system to exhibit a single mode with   $\pi$ quasienergy at each end of the chain. In particular we set  $\Delta=w=1$, $\bar{\mu}=0.6$  and the time period is taken to be $T=8.25$.  These parameters are chosen to match those of Ref.~\cite{Yates_2019} but the results can easily be generalized.  Figure~\ref{fig:Ainf}-a shows $A_\infty$ for our system for a single realization of disorder and a range of disorder strengths. The figure shows that as the disorder increases the autocorrelation  becomes  scattered.  Figs.~\ref{fig:Ainf}-b and ~\ref{fig:Ainf}-c portray the tight binding Krylov/Arnoldi model.  The panel c represents the matrix elements of $i\ln(W)$, which is the effective Arnoldi Hamiltonian, by color, where darker color corresponds to larger magnitude.  It shows that the hopping along the chain is dominated by  nearest neighbor bonds corresponding to the two diagonals adjacent to the main one.  All other matrix elements are strongly suppressed.  We denote the nearest neighbor hopping between sites $m$ and $m+1$ by $\beta_m$ and present them in ~\ref{fig:Ainf}-b. We find that $\beta_m$ oscillates along the chain.  This resembles the Su–Schrieffer–Heeger  model \cite{SSH} and is in agreement with a single edge mode.  Indeed, for this particular set of parameters and in the clean limit, the system is in the $\pi$-phase and therefore one would expect it to map to a tight binding model with a single Majorana mode on each edge. In the clean case, $A_\infty$ is indicative of a strong mode which means $A_\infty$ plateaus at a non-zero value for a long time but eventually decays to zero (not shown in the plot) at a time which is exponential in the system size. When the disorder is turned on, both $A_\infty$ and the nearest neighbor hopping change their character, and when the disorder is high the autocorrelation becomes very noisy.  
		
\begin{figure}
		\centering
		\includegraphics[width=1.0\columnwidth]{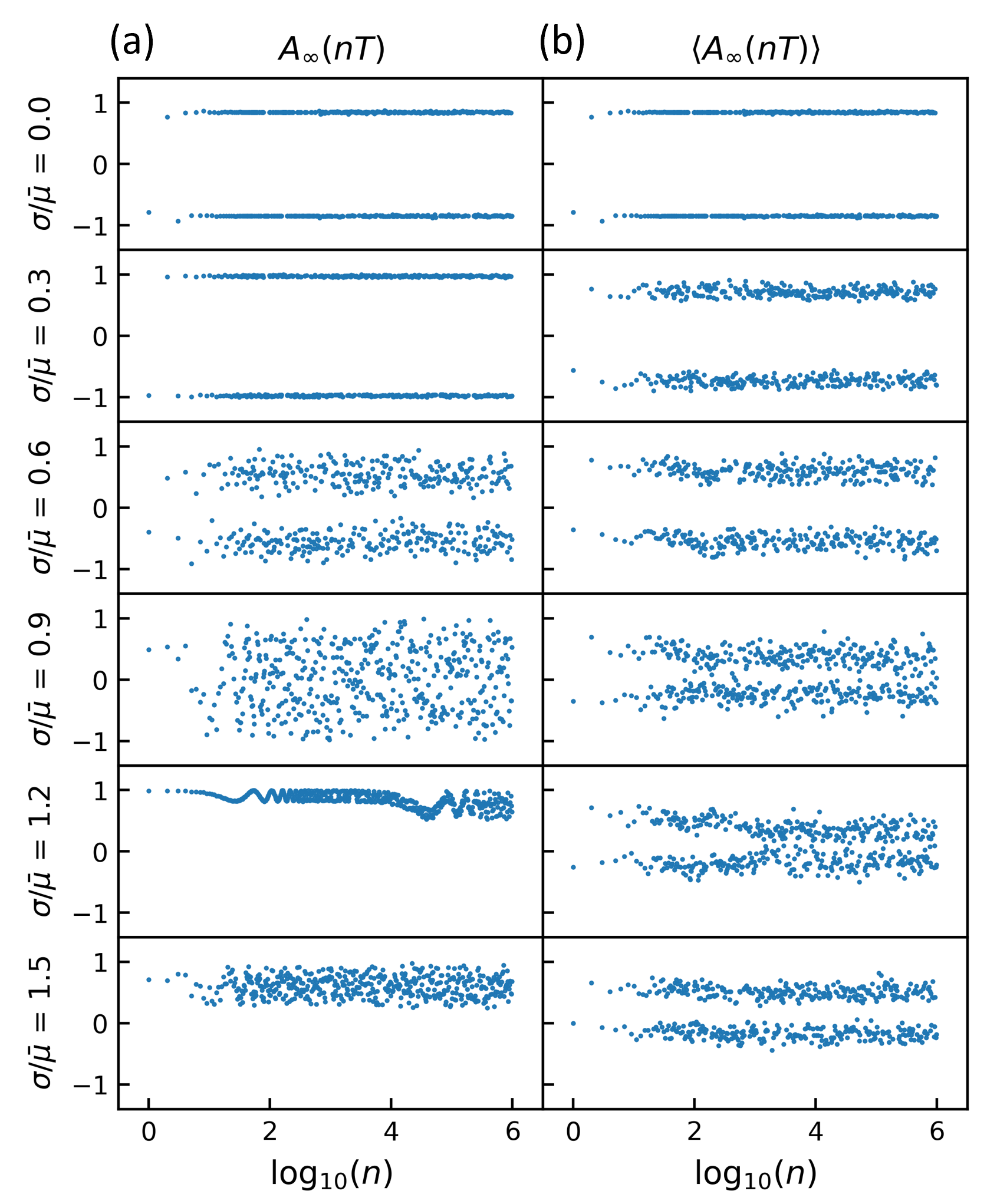}
		\caption{(a) A typical autocorrelation function $A_\infty(nT)$ as a function of $\log_{10}(n)$, calculated using exact diagonalization of a chain of $L=200$ sites, shown for varying disorder strength. (b) The same data as on the left, but averaged over 10 disorder realizations. 
			The model parameters used are the same as in Fig.~\ref{fig:Ainf}.}
		\label{fig:A_inf}
	\end{figure}

	Taking a closer look at $A_\infty$  of the disordered chain one can identify signatures of a phase transition.  In the clean limit, when a $\pi$-Floquet Majorana mode is present, the autocorrelation oscillates with strobosocopic time as $(-1)^n$, and eventually decays to zero due to finite system size \cite{Yates_2019}. The behavior of $A_\infty$ for a particular disorder realization is shown  in Fig.~\ref{fig:A_inf}.  
	In panel~\ref{fig:A_inf}-b, an average over several disorder profiles is shown.  One can see a hint of a phase transition as the average autocorrelation goes from oscillating symmetrically about zero to oscillating symmetrically about a non-zero value, as the disorder strength is increased.  
	
\begin{figure}
		\centering
		\includegraphics[width=1.0\columnwidth]{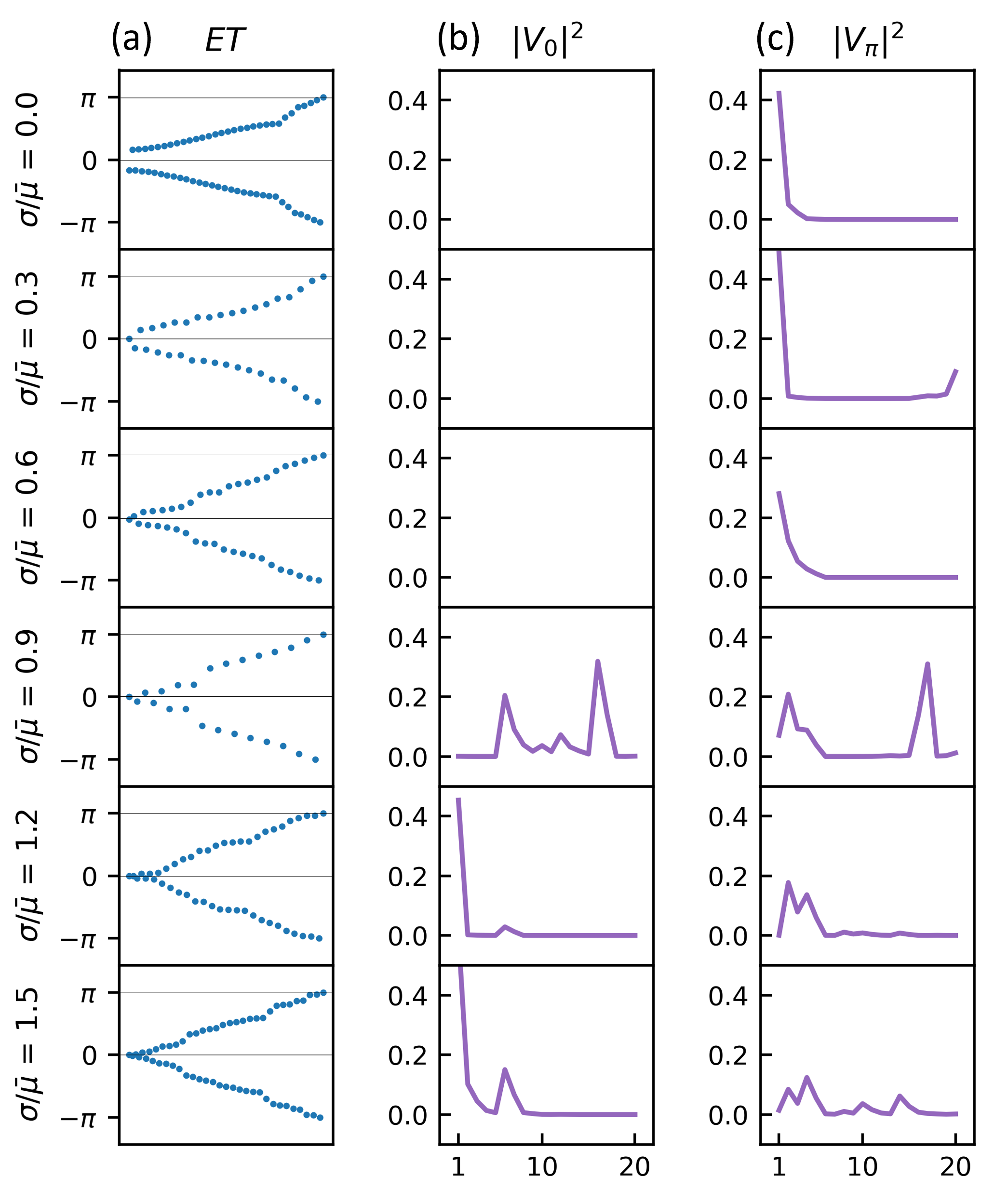}
		\caption{Arnoldi spectrum, i.e., spectrum of $i\ln(W)$, and wave functions, for a single disorder realization.  (a) The spectrum of the Arnoldi chain plotted in order of increasing quasienergy magnitude (b) Probability distribution for the state at zero energy. The first 20 Arnoldi sites are shown. (c) Probability distribution for the state at energy $\pi$. Besides the range of disorder strengths, the parameters are the same as in Fig.~\ref{fig:Ainf}. In some cases, there is 
 a state near zero energy in the Arnoldi spectrum which is not localized at the edge of the sample.  In that case it is not plotted in the middle panel.}
		\label{fig:Arnoldi_spectrum}
	\end{figure}

	Despite the noisy $A_\infty$, and the disordered nearest neighbor hopping amplitude, the Krylov space is still indicative of strong edge modes.  First, the agreement between exact diagonalization calculation of $A_\infty$ and the Krylov calculation remains  good even in the presence of disorder.  While the mapping onto the Krylov space preserves the topology, our method requires the truncation of the space and therefore the agreement with exact diagonalization is reassuring. Further insight can be gained by diagonalizing the Krylov tight binding model. To this end we  diagonalize the matrix $W$ whose logarithm gives the energies of the Krylov model.  Fig.~\ref{fig:Arnoldi_spectrum} shows the logarithm of the spectrum of W for various disorder strengths.  At low disorder strengths there are two $\pi$ energy states in the physical system.  In the Krylov space we are only concerned with the left edge of the system (due to the choice of seed operator), and we therefore see one $\pi$-mode eigenstate of $W$.  Remarkably, when the disorder is increased, an additional Majorana mode 
	appears, associated with zero quasienergy.  The zero and $\pi$ eigenstates of $W$ are plotted in columns b and c of Fig.~\ref{fig:Arnoldi_spectrum} where approximate zero and $\pi$ modes are shown, when they exist.
	
\begin{figure}
		\centering
		\includegraphics[width=1.0\columnwidth]{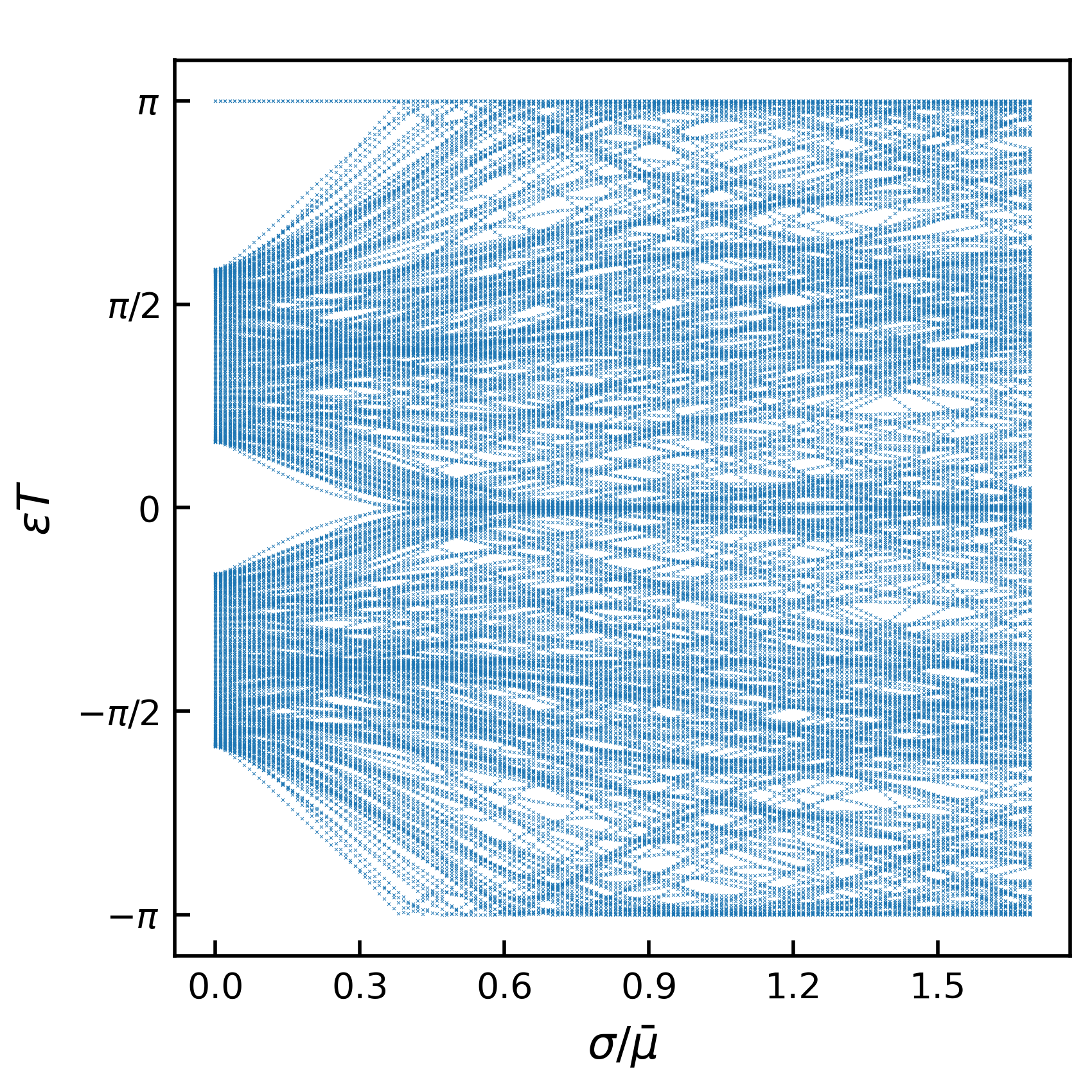}
		\caption{Evolution of the spectrum as the disorder strength is increased. As the disorder is increased the quasienergy gap around $\epsilon=\pi$ closes but the the edge MPMs remain localized.  Moreover, the gap around $\epsilon=0$ also closes and a MZM emerges.  The Majorana edge modes can be seen in Fig~\ref{fig:ED_wavefunctions} }
		\label{fig:ED_spectrum}
	\end{figure}

\begin{figure}
		\centering
		\includegraphics[width=1.0\columnwidth]{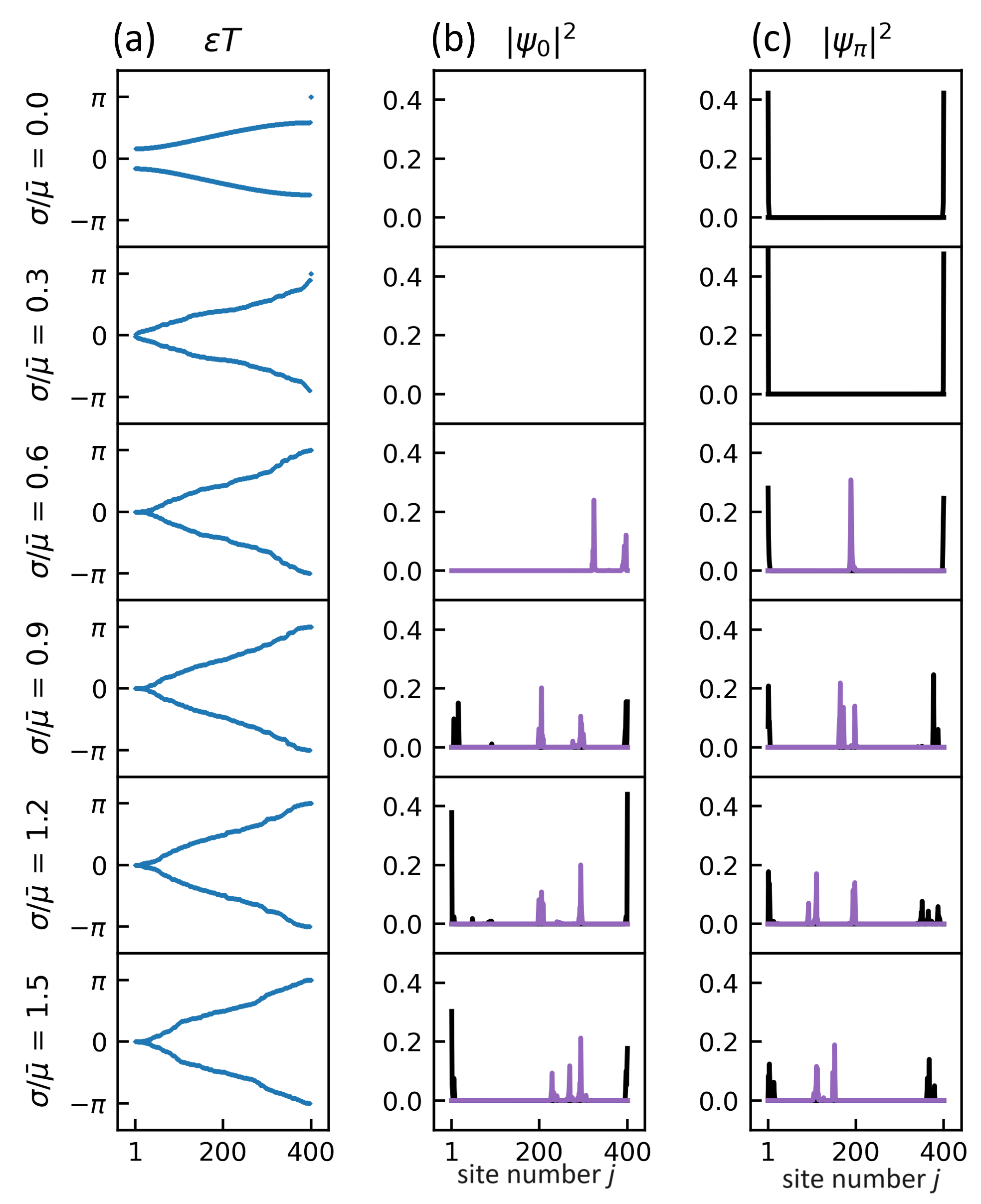}
		\caption{Exact diagonalization spectrum and wavefunctions for a single disorder realization with increasing disorder strength (top to bottom). (a) Spectrum, (b) zero modes,  (c) $\pi$-modes. In both  the middle and right columns the black curves are states at energies indistinguishable from $\epsilon T=0$ or $\epsilon T =\pi$.  The purple curves are states with energies in the vicinity of zero or $\pi$.}
		\label{fig:ED_wavefunctions}
	\end{figure}

	We now compare the above results with those of exact diagonalization. Fig.~\ref{fig:ED_spectrum} shows the spectrum of the physical system, obtained from exact diagonalization, for one realization of disorder, while the corresponding states are plotted in Fig.~\ref{fig:ED_wavefunctions}. Fig.~\ref{fig:ED_spectrum} shows that as the disorder is increased, the gaps in the spectrum around $\epsilon T=\pi,0$ close.  While the gaps do not reopen, the Majorana modes that initially exist at $\pi$ still persist, as can be seen in Fig.~\ref{fig:ED_wavefunctions}-c.  Moreover, as the gap closes at zero quasienergy, new Majorana modes appear at zero quasienergy, with their wavefunctions depicted in Fig.~\ref{fig:ED_wavefunctions}-b.  The effect where despite gap closings, the edge modes in the gaps stay localized, can be attributed to the localization of the bulk modes. Thus there is no bulk channel available for the edge modes on the two ends of the chain to hybridize.    
    In Figs~\ref{fig:ED_wavefunctions}-b,c, edge MZMs and MPMs are plotted in black while other states around the same energy (if they exist) are plotted in purple.
 These additional states are also localized but are usually pinned to the disorder away from the edge. 
 
 Note that in the disordered case, the spectrum is not a useful tool for identifying strong modes as the gaps are closed.  We therefore extend our insight by looking in Krylov space, and also by calculating the topological entanglement entropy (TEE). The latter is presented in the next section. 
	
	\section{Topological entanglement entropy}\label{sec:TEE}
	The exact diagonalization and Arnoldi analysis point to a disorder induced phase transition in the driven Majorana chain. Surprisingly, the transition out of the $\pi$-phase is not to a trivial phase but to a richer topological phase where both MZMs and MPMs exist.
 We see an indication for the transition in Arnoldi space as new zero modes appear in the spectrum of the Arnoldi chain. These modes also appear in the spectrum of the original Kitaev chain for a single disorder realization.  Before turning on the disorder, the clean system is in the Floquet Majorana $\pi$ phase and exhibits $\pi$ Majorana modes.  Once the disorder is turned on, and increased in strength, all of the states in the spectrum become progressively localized and the energy gaps around zero and $\pi$ quasienergies close.  For large enough disorder, additional Majorana modes appear around zero quasienergy which suggests that the system has transitioned into  the $\pi$-$0$ phase. 
	However, in the  disordered case, quasienergy states at zero/$\pi$ may appear accidentally and do not necessarily imply topological protection.  
	
\begin{figure}
		\centering
		\includegraphics[width=\columnwidth]{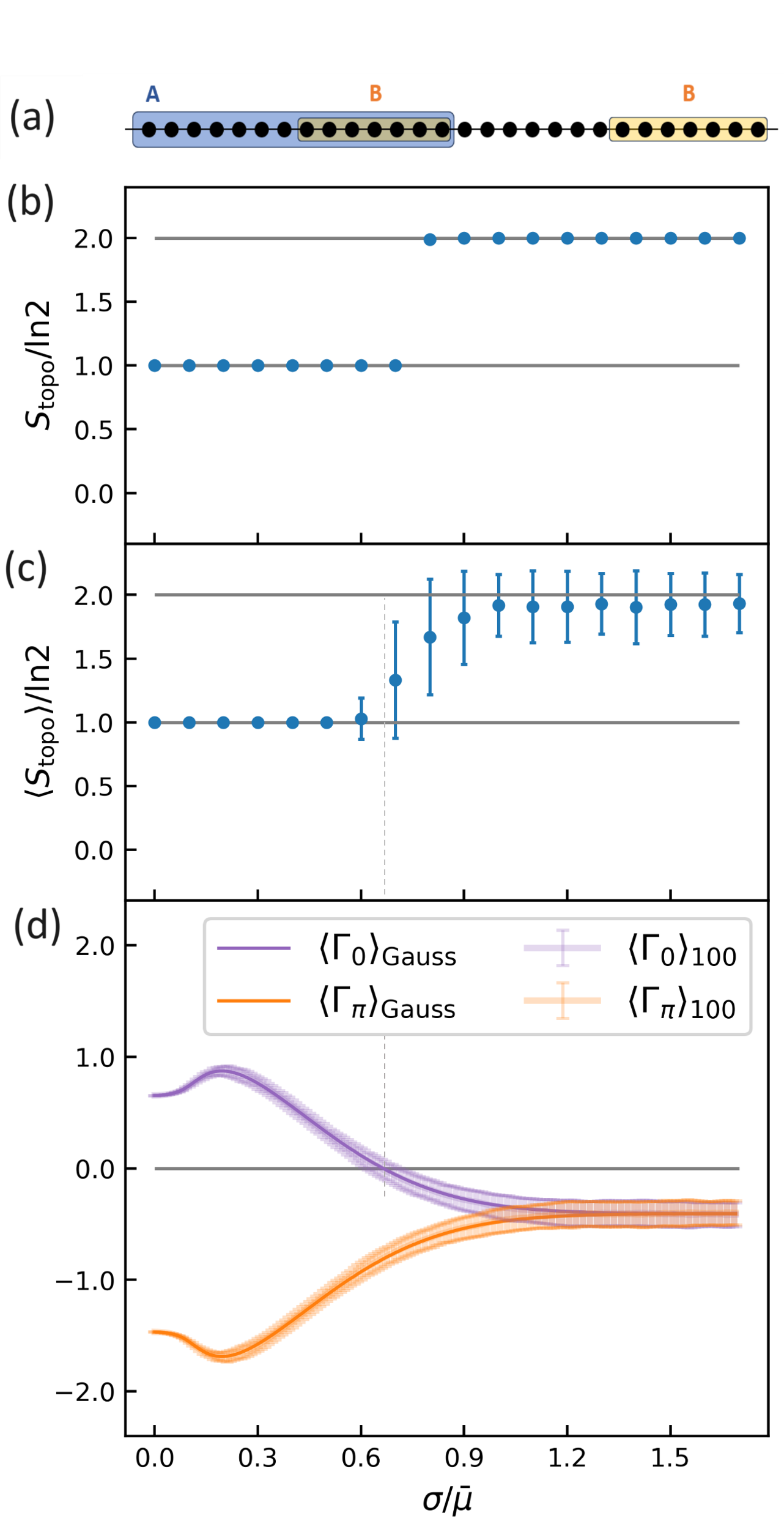}
		\caption{The topological entanglement entropy and inverse localization length. (a) Partitions used to calculate the topological entanglement entropy.  (b-c) The topological entanglement entropy as a function of the disorder strength for a single disorder realization (b), and averaged over 100 disorder profiles (c).  In both cases a chain of 200 sites was used.
  (d) The spatial decay rates $\Gamma_0$ and $\Gamma_\pi$ (inverse localization length), in units of inverse lattice constant, corresponding to the $0$ and $\pi$ modes respectively. The light (orange)purple lines are ($\Gamma_\pi$)$\Gamma_0$, calculated  using Eqns.~\ref{eq:gamma0} (\ref{eq:gammaPi}) averaged over 100 disorder realizations with error bars corresponding to the standard deviation. The average of these 100 realizations exactly coincide with (Eqs.~\ref{eq:gammaPiDisorder}),\ref{eq:gamma0Disorder}, ploted in darker shades, where the average is done by integrating a Gaussian distribution for $\delta\mu$. Note that the condition for an edge mode is $\Gamma < 0$ as the opposite regime represents a non-normalizable function.  In both panels (c) and (d) a phase transition is observed around $\sigma/\bar\mu \approx 0.6$, as indicated by the vertical dashed line. }
		\label{fig:TEE}
	\end{figure}

	To address this question, we recall that the  entanglement entropy contains a subleading term which characterizes global features of a quantum state \cite{Preskill06,Levin2006}. In order to separate it from other subleading terms  which may be sensitive to the disorder, we calculate the TEE\cite{Zeng2015,Zeng_2016,Fromholz_2020} which is a difference of entanglement entropies of different partitions, such that one of them is a disconnected partition\cite{Casini_2004}.  Fig.~\ref{fig:TEE}-(a) shows an example of possible partitions.  In this example partition $A$ is connected while partition $B$ is disconnected and the TEE is given by:
	\begin{eqnarray}
		S^{\rm topo} = S_A+S_B-S_{A\cup B}-S_{A\cap B}.
	\end{eqnarray}
	
	In equilibrium, it  was found that the TEE gives $\log(2)$ in the topological phase and zero in the trivial phase of the Kitaev chain.  This value is robust to disorder \cite{Levy_2019} and interactions 
 \cite{Fromholz_2020}.  In the driven case we have four phases \cite{Khemani_2016} and two corresponding $Z_2$ invariants.  We therefore expect that in the $\pi$ phase the TEE will be $\ln(2)$, and  $2\ln(2)$ in the $\pi$-$0$ phase. The observation of a jump from $\ln(2)$ to $2\ln(2)$ is therefore consistent with a disorder induced transition between the $\pi$ phase and the $\pi$-$0$ phase. While different disorder realizations transition at different disorder strengths, the average over many realizations shows a clear phase transition. 
	Fig.~\ref{fig:TEE}-b shows the TEE as a function of disorder strength for one representative disorder profile and a disorder averaged TEE is shown in Fig~\ref{fig:TEE}-c.  
	
	The calculation of the TEE follows Ref.~\cite{Fromholz_2020} where we use the eigenstates of $\ln(U(T))$ to build the density matrix.  We note that the calculation works best on systems without boundaries since edge Majoranas lead to degeneracies, causing a confusion as to which state to include in the density matrix.  We therefore add a small hopping and pairing amplitude between the last site of the chain and its first, which slightly lifts this degeneracy. 
	
	\section{Exact solution for zero and $\pi$ modes for a disordered semi-infinite chain} \label{sec:exact}
	Another approach that may shed light on the effect of disorder on our system is to study the semi-infinite chain where the edge MZMs and MPMs can be found exactly.  When the chain is semi-infinite any edge modes are completely decoupled from their couterparts (which are an infinite distance away) and are therefore exactly at quasienergy zero or $\pi$. For a given disorder profile, we can therefore use the transfer matrix method to find the state and see whether it is indeed localized.  
	
	For this purpose we express the zero/$ \pi$ Majorana modes as a linear combination 
	of Majorana operators on all sites:  
	\begin{eqnarray}
		\Psi_{0/\pi}=\sum_{j}\left(a_j\gamma_{aj}+b_j\gamma_{bj}\right),
	\end{eqnarray}
	where $\gamma_{aj}$ and $\gamma_{bj}$ are the Majoranas corresponding to the real and imaginary parts of the fermion operator on site $j$, and $a_j$, $b_j$ are scalar coefficients.  The  $\Psi_{0/\pi}$ operators satisfy the following time evolution  over a single cycle: 
	\begin{eqnarray}
		U(T)\Psi_0 U^\dagger(T)= \Psi_0, \nonumber \\
		U(T)\Psi_\pi U^\dagger(T)=-\Psi_\pi.
	\end{eqnarray}
	This allows us to obtain a recursive relation for the coefficients $a_j$ and $b_j$ for both modes. We first write the Hamiltonians in Eq.~\eqref{eq:Hamiltonians} in terms of the Majorana operators:
	\begin{eqnarray}
		{\cal H}_1(w=\Delta) &=& -i\Delta\sum_j\gamma_{aj+1}\gamma_{bj}, \nonumber \\
		{\cal H}_2 &=& -i\sum_j{\mu_j\over 2}\gamma_{aj}\gamma_{bj}.
	\end{eqnarray}
	It follows that,
	\begin{eqnarray}
		\nonumber
		& U\gamma_{aj}U^\dagger=\\
		&C_\Delta C_j \gamma_{aj}+C_\Delta S_j \gamma_{bj} + S_\Delta C_{j-1} \gamma_{bj-1} - S_\Delta S_{j-1} \gamma_{aj-1},\nonumber \\
		\nonumber
		& U\gamma_{bj}U^\dagger = \\
		&C_\Delta C_j \gamma_{bj}-C_\Delta S_j \gamma_{aj} - S_\Delta C_{j+1} \gamma_{aj+1} - S_\Delta S_{j+1} \gamma_{bj+1},\nonumber \\
	\end{eqnarray}
	where $C_\Delta = \cos{\Delta T}$, $S_\Delta = \sin{\Delta T}$ and $C_j = \cos{\mu_j T/2}$, $S_j = \sin{\mu_j T/2}$. Using these relation we can find the recursion relation for the coefficients $a_j$ and $b_j$ for $j\geq 2 $:
	\begin{eqnarray}
		\nonumber
		&e^{-i\epsilon T}  b_j = \\
		&C_\Delta C_j b_j +C_\Delta S_j a_{j} +S_\Delta C_{j} a_{j+1} - S_\Delta S_{j} b_{j-1},\nonumber \\
		\nonumber
		& e^{-i\epsilon T} a_{j}=\\
		&C_\Delta C_{j} a_{j}-C_\Delta S_{j} b_{j} - S_\Delta C_{j} b_{j-1} - S_\Delta S_{j} a_{j+1},\nonumber \\
	\end{eqnarray}
	where $\epsilon T =0$ or $\epsilon T = \pi$. The boundary conditions are derived from the operation of  $U(T) $  at the end of the chain. The resulting equations read:
	\begin{eqnarray}\label{eq:BC_Tmtx}
		\nonumber
		&e^{-i\epsilon T}  b_1 = C_\Delta C_1 b_1 + S_1 a_{1} +S_\Delta C_{1} a_{2},\nonumber \\
		& e^{-i\epsilon T} a_{1}= C_{1} a_{1}-C_\Delta S_{1} b_{1} - S_\Delta S_{1} a_{2}. 
	\end{eqnarray}
	We can write the recursion relation for  $j\geq 2 $ as a matrix recursion relation:
					\begin{align}&
							\nonumber
							\left(\begin{array}{cc}
								S_\Delta C_{j}  &C_\Delta C_{j}  -e^{-i\epsilon T} \\
								S_\Delta S_{j}& C_\Delta S_{j}
							\end{array}\right)\left(\begin{array}{c}
								a_{j+1}\\
								b_{j}
							\end{array}\right)=\\
							&\left(\begin{array}{cc}
								-C_\Delta S_j &S_\Delta S_j  \\
								-e^{-i\epsilon T}+C_\Delta C_j & -S_\Delta C_j
							\end{array}\right)\left(\begin{array}{c}
								a_{j}\\
								b_{j-1}
							\end{array}\right)
						\end{align}
						and invert it to obtain the transfer matrix $\tau_j(\epsilon)$ for $j\geq 2$:
						\begin{align}
							\left(\begin{array}{c}
								a_{j+1}\\
								b_{j}
							\end{array}\right)= \tau_j(\epsilon)
							\left(\begin{array}{c}
								a_{j}\\
								b_{j-1}\end{array}\right),
						\end{align}
						with 
						\begin{align}
							\!\!\!\!\!\!\!\!&\tau_j(\epsilon) =\frac{1}{S_j}
							\left(
							\begin{array}{cc}
								2 \frac{C_\Delta}{S_\Delta} C_j- \frac{e^{-i \epsilon T }}{S_\Delta}-e^{i \epsilon T }  \frac{C_\Delta^2}{S_\Delta} & e^{i \epsilon T } C_\Delta  -C_j \\
								e^{i \epsilon T } C_\Delta  -C_j& -e^{i \epsilon T } S_\Delta \\
							\end{array}
							\right).
						\end{align}

\begin{figure}
							\centering\includegraphics[width=0.9\columnwidth]{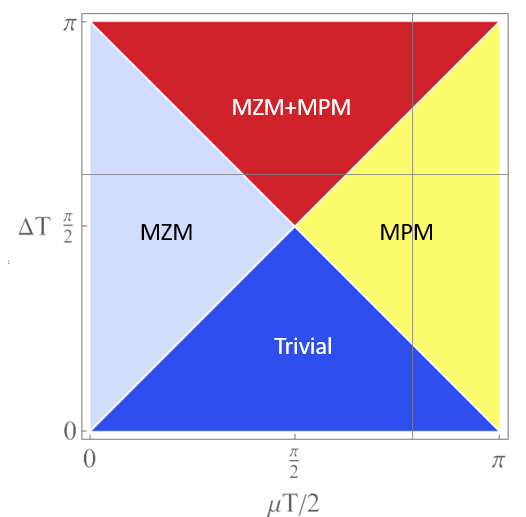}
							\caption{The phase diagram for a clean system based on the criteria \eqref{eq:gamma0} and \eqref{eq:gammaPi}. Here blue is the trivial phase, light blue is the phase with a MZM, yellow has a MPM and red is a MZM+MPM phase. The grid lines indicate the parameters chosen for much of this paper, $\mu =0.6$ and $\Delta T \mod 2\pi $ with $\Delta =1 $ and $T=8.25 $.}
							\label{fig:cleanPD}
						\end{figure}

						In order to find the wave function of the zero/$\pi $ Majorana modes, localized on the  edge of a  semi-infinite wire, we take the first coefficients to be  $(a_1,b_0)^T = \left(1,0 \right)^T$, then apply the boundary recursion relation \eqref{eq:BC_Tmtx}, followed by  repeated application of $\tau_{j}(\epsilon)$. 
						We calculate the resulting state entries at some large distance $(a_{N+1},b_N)^T $. We find that $(a_{N+1},b_N)^T\sim e^{N\Gamma_0}(1,\tan(\Delta T/2))^T $, with the inverse localizations length $\Gamma_{0/\pi}$ given by: 
						\begin{align}\label{eq:gamma0}
							\Gamma_0 &=\frac{1}{N}\ln\left[\cot^N\left(\frac{\Delta T }{2}\right) \prod_{j=1}^N \tan\left(\frac{\mu_j T}{4}\right)\right] \nonumber\\
							&=\ln\left[ \cot\left(\frac{\Delta T}{2}\right)\right]+{1\over N}\sum_{j=1}^N \ln\left[\tan\left(\frac{\mu_j T}{4}\right)\right].
						\end{align}
						For this to be normalizable we require: 
						${\rm Re}\left(\Gamma_0<0 \right)$.
						The criteria for the $\pi $ mode can be determined similarly. We find:
						$(a_{N+1},b_{N})^T \sim e^{N\Gamma_\pi}(1,-\tan(\Delta T/2))^T $
						with:
						\begin{equation}\label{eq:gammaPi}
							\Gamma_\pi 
							=\ln\left[ \cot\left(\frac{\Delta T}{2}\right)\right]+{1\over N}\sum_{j=1}^N \ln\left[\cot\left(\frac{\mu_j T}{4}\right)\right].
						\end{equation}
						For a clean system where $\mu_j=\bar \mu$ these two criteria produce the phase diagram in Fig. \ref{fig:cleanPD}, and are in agreement with the localization lengths calculated in \cite{Yates_2019}.
						
\begin{figure}
							\centering
							\includegraphics[width=0.9\columnwidth]{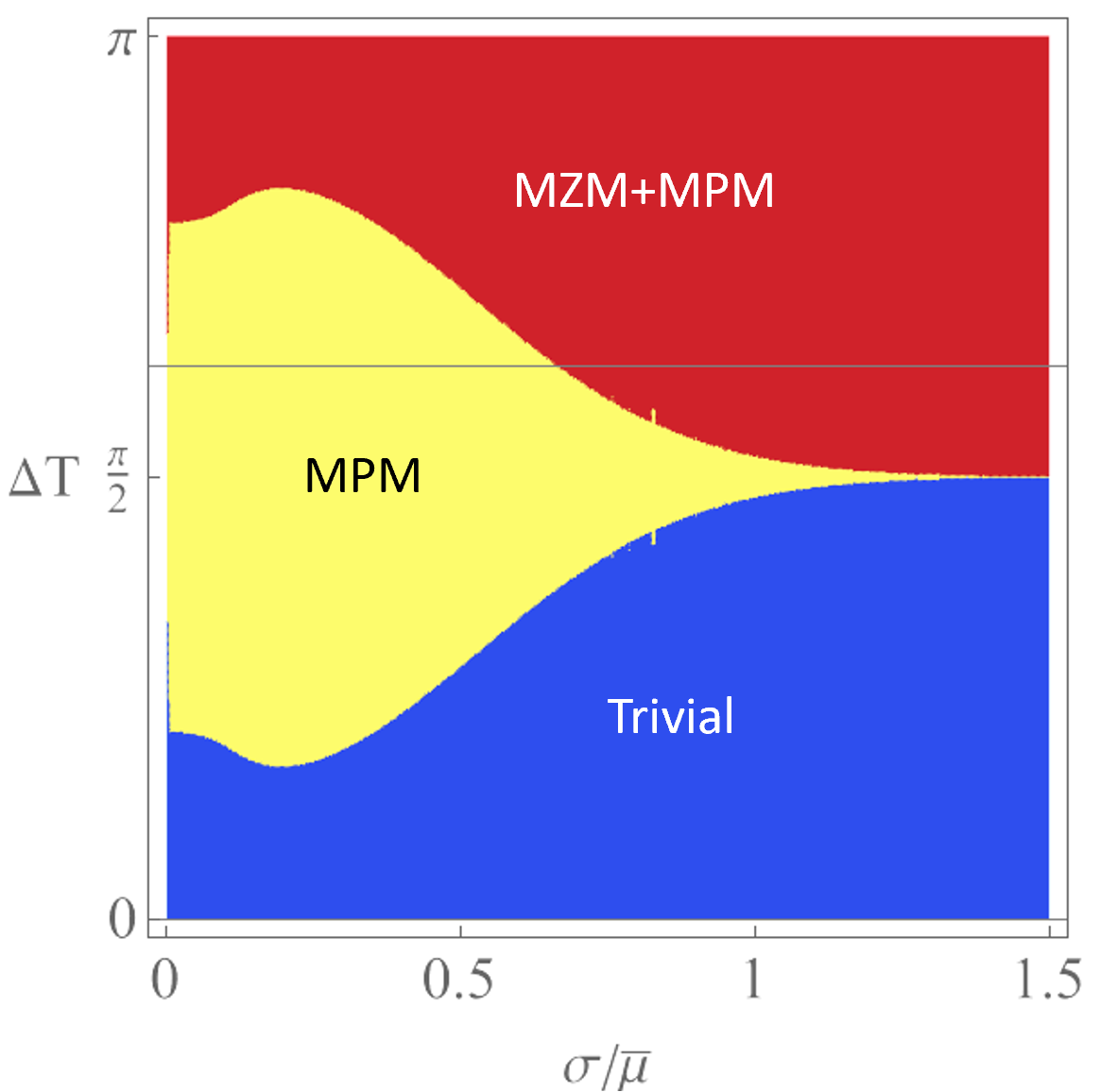}
							\caption{The phase diagram for a disordered system with $ \bar \mu=0.6$ and $T=8.25 $  based on the criteria \eqref{eq:gamma0Disorder} and \eqref{eq:gammaPiDisorder} where $ \delta \mu_i$ is taken from a Gaussian distribution with variance $\sigma^2 $. Here blue is the trivial phase, yellow is the phase with a MPM and red is a MZM+MPM phase.  The grid line indicates $\Delta T \mod 2\pi $ with $\Delta =1 $ and $T=8.25$, as chosen for much of this paper.}
							\label{fig:disorderedPD1}
						\end{figure}

\begin{figure}[t]
							\centering
							\includegraphics[width=\columnwidth]{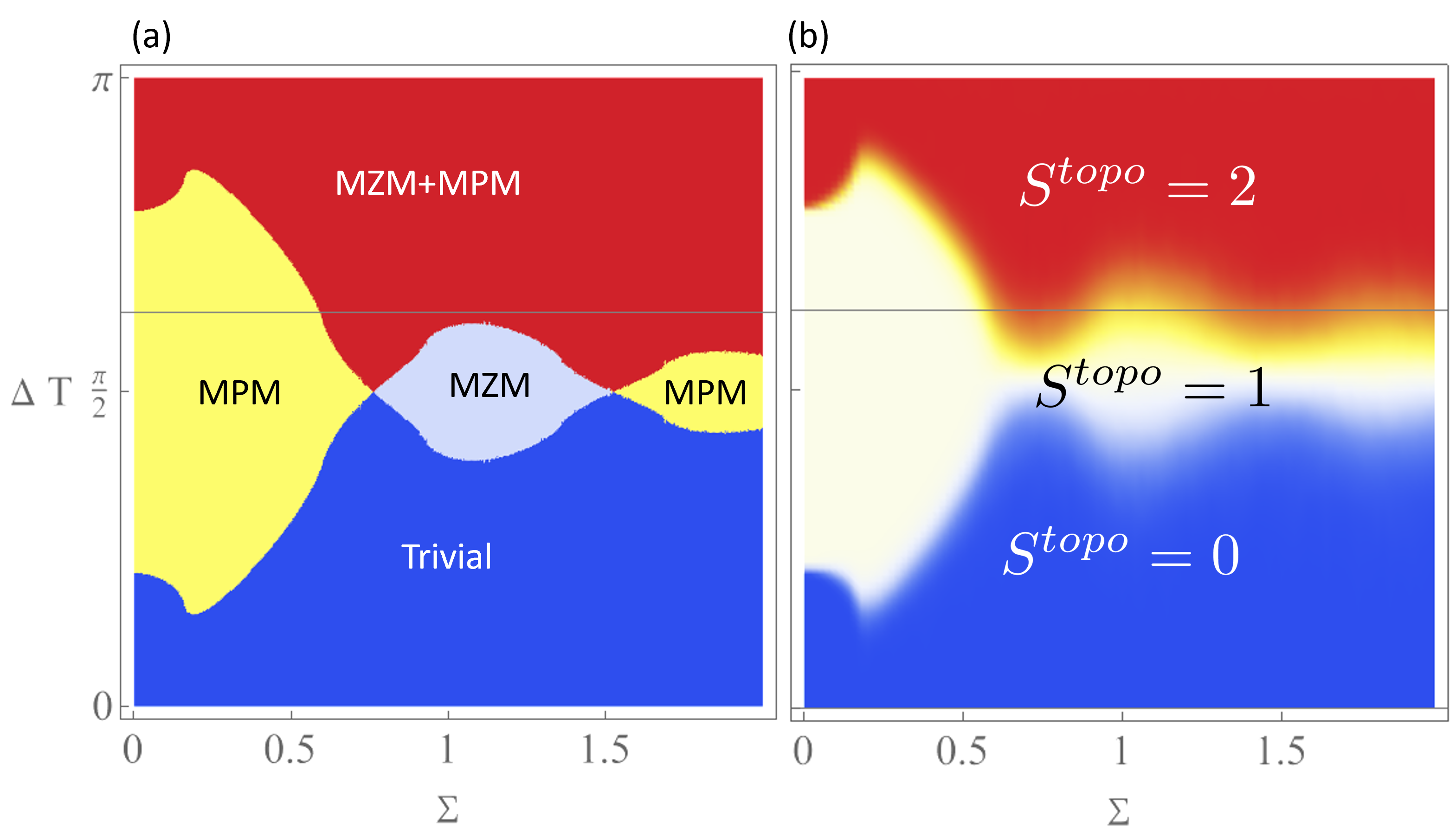}
							\caption{(a) The phase diagram for a disordered system with $ \bar \mu=0.6$ and $T=8.25 $  based on the criteria \eqref{eq:gamma0Disorder} and \eqref{eq:gammaPiDisorder} for a uniform distribution in the range $[-\Sigma/2,\Sigma/2] $. Here blue is the trivial phase, light blue is the phase with a MZM, yellow has a MPM, and red is the MZM+MPM phase.  (b) The topological entanglement entropy, $S^{\rm topo}$, calculated for the same parameters as on the left panel.  The chain length is $L=100$ and the results are averaged over 1000 disorder realizations. The grid line in both figures indicates $\Delta T \mod 2\pi $ with $\Delta =1 $ and $T=8.25$, as chosen for much of this paper}. 
							\label{fig:disorderedPD2}
						\end{figure}

						In the disordered case we may perform a disorder average over many realizations of disorder of a known distribution: 
						\begin{align}\label{eq:gamma0Disorder}
							\bar{\Gamma}_0 
							&=\ln\left[ \cot\left(\frac{\Delta T}{2}\right)\right]+\left \langle \ln\left[\tan\left({(\bar{\mu}+\delta \mu)T\over 4}\right)\right]\right\rangle,
						\end{align}\begin{align}\label{eq:gammaPiDisorder}
							\bar{\Gamma}_\pi 
							&=\ln\left[ \cot\left(\frac{\Delta T}{2}\right)\right]+\left \langle \ln\left[\cot\left(\frac{(\bar{\mu}+\delta \mu)T}{4}\right)\right]\right\rangle,
						\end{align}
						where $ \langle \cdot  \rangle $ represents an average over the  random variable $\delta \mu_i $.
						Fixing the average chemical potential to $\bar \mu=0.6 $ and assuming  $\delta \mu_i $ is taken from a Gaussian distribution with zero mean and $ \sigma^2 $ variance, these two criteria produce the disordered phase diagram \ref{fig:disorderedPD1}. If instead $ \delta \mu_i$ is chosen with a uniform distribution from the range $[-\Sigma/2,\Sigma/2] $,  one obtains the disordered phase diagram shown in Fig.~\ref{fig:disorderedPD2}.  For comparison, we have also calculated the TEE for the same parameters.  We used exact diagonalization and averaged the TEE over 1000 realizations of disorder at each point.  The TEE cannot distinguish between the $\pi$ and zero mode phases, as they both correspond to $S^{\rm topo} =\ln(2)$. But the TEE does identify the phases where two topological edge modes exist simultaneously by giving $S^{\rm topo} =2\ln(2)$. 
      The phase transition from MPM to the MPM-MZM phase, as obtained from the TEE,  is in agreement with that obtained using $\Gamma_0$ and $\Gamma_{\pi}$, see Fig~\ref{fig:TEE} (d).
						
						At low disorder we can expand the above expressions for $\Gamma_0$ and $\Gamma_\pi$ to see in which direction the disorder is changing them:
						\begin{eqnarray}
							\bar{\Gamma}_0 \approx \ln\left[{\cot\left(\frac{\Delta T}{2}\right)\tan\left(\frac{\bar \mu T}{4}\right)}\right] -{T^2\over 8}{\cot\left({\bar \mu T}\right)\over \sin\left({\bar \mu T}\right)} \sigma^2, \nonumber \\
							\bar{\Gamma}_\pi \approx \ln\left[{\cot\left(\frac{\Delta T}{2}\right)\cot\left(\frac{\bar \mu T}{4}\right)}\right] +{T^2\over 8}{\cot\left({\bar \mu T}\right)\over \sin\left({\bar \mu T}\right)} \sigma^2.
						\end{eqnarray}
						Above $\sigma^2=\langle\delta\mu^2\rangle$ is the variance of the disorder.  It can be seen from above that while $\Gamma_\pi$ is increased by weak disorder (i.e, the $\pi$ Majorana modes become less localized), $\Gamma_0$ is decreased by disorder, and therefore  the zero modes become more localized.
      Fig.~\ref{fig:TEE}-d shows the decay length of the edge modes.  The solid purple and orange lines represent the decay length calculated using Eqs.~\eqref{eq:gamma0}, \eqref{eq:gammaPi} averaged over a 100 realizations of disorder.  These lines exactly coincide with $\Gamma_{0/\pi}$ obtained by assuming a Gaussian distribution of $\delta\mu$ and performing the average by integration of Eq.~\eqref{eq:gamma0Disorder} and
       (Eq.~\eqref{eq:gammaPiDisorder}).
      
    \section{Conclusions}\label{sec:Conclusions}
	We have studied the effect of disorder on a driven Kitaev chain.  
      We approach the study of the disordered system by mapping it on to a Krylov space using the Arnoldi method.  This mapping results in an effective tight binding chain whose edge states can be linked to the topology of the system.  We complement this analysis with exact diagonalization and find good agreement between the two methods.  
						
		We find that the disorder may induce a topological phase transition. For example, when starting from a clean system in the $\pi$ phase (with only MPMs on the boundary), increasing the disorder causes a transition to the $\pi$-$0$ phase.  In this new phase each edge of the sample hosts  a $\pi$-Majorana mode as well as a Majorana mode at zero quasienergy.  This transition occurs at a relatively high disorder where a mapping to an equivalent clean system is not possible. 
        The phenomenon of emergent or strengthened edge modes as a result of disorder can be understood in the context of localization.  Disorder may cause localization of bulk modes around impurities (or disorder pinning sites), generally far from the edges.  The localized modes therefore have very little overlap with the edge Majoranas and do not cause hybridization.  Moreover, the localized states signify the opening of a mobility gap which can enable new edge mode formation.  This is similar to the phenomenon of Anderson topological insulator \cite{Li_2009} and Anderson topological superconductor \cite{Borchmann_2016} reported earlier.
                        
     Since characterizing the system through its spectrum is complicated because of the closing of the energy gap, we turn to calculating the TEE.  The TEE shows a clear transition between two values: $\ln(2)$ corresponding to a single pair of Majorana $ \pi$ modes at low disorder, and $2\ln(2)$ corresponding to two pairs of Majorana modes at high disorder, one at quasienergy $\epsilon T = 0$ and the other at quasienergy $\epsilon T = \pi$. The topological phase transition is further confirmed by an exact computation of the localization lengths of the edge modes by a transfer matrix method.  The exact solutions allow us to find the topological phase transition for every disorder profile as well as the disorder averaged transition.  Moreover, the extracted localization lengths of the Majorana modes provide an insight into the robustness of the modes with respect to disorder.

						\begin{acknowledgments}
							We would like to acknowledge useful discussions with Graham Kells.  HL would like to thank Joseph Ming Tak Ling for continual support and encouragement, and key lessons in mathematics.  TPB acknowledges financial support from NSERC and INTRIQ.  This work was supported by the US Department of Energy, Office of
							Science, Basic Energy Sciences, under Award No.~DE-SC0010821 (AM), the National Science Foundation (NSF) through award numbers MPS-2228725 and DMR-1945529 and the Welch Foundation through award number AT-2036-20200401 (MK and SRK) and has benefited from the hospitality of the Aspen Center for Physics which is supported by National Science Foundation grant PHY-2210452 (DM,MK,TPB).
						\end{acknowledgments}
						
						\bibliography{FloquetKitaev.bib}

\begin{thebibliography}{69}%
\makeatletter
\providecommand \@ifxundefined [1]{%
 \@ifx{#1\undefined}
}%
\providecommand \@ifnum [1]{%
 \ifnum #1\expandafter \@firstoftwo
 \else \expandafter \@secondoftwo
 \fi
}%
\providecommand \@ifx [1]{%
 \ifx #1\expandafter \@firstoftwo
 \else \expandafter \@secondoftwo
 \fi
}%
\providecommand \natexlab [1]{#1}%
\providecommand \enquote  [1]{``#1''}%
\providecommand \bibnamefont  [1]{#1}%
\providecommand \bibfnamefont [1]{#1}%
\providecommand \citenamefont [1]{#1}%
\providecommand \href@noop [0]{\@secondoftwo}%
\providecommand \href [0]{\begingroup \@sanitize@url \@href}%
\providecommand \@href[1]{\@@startlink{#1}\@@href}%
\providecommand \@@href[1]{\endgroup#1\@@endlink}%
\providecommand \@sanitize@url [0]{\catcode `\\12\catcode `\$12\catcode
  `\&12\catcode `\#12\catcode `\^12\catcode `\_12\catcode `\%12\relax}%
\providecommand \@@startlink[1]{}%
\providecommand \@@endlink[0]{}%
\providecommand \url  [0]{\begingroup\@sanitize@url \@url }%
\providecommand \@url [1]{\endgroup\@href {#1}{\urlprefix }}%
\providecommand \urlprefix  [0]{URL }%
\providecommand \Eprint [0]{\href }%
\providecommand \doibase [0]{https://doi.org/}%
\providecommand \selectlanguage [0]{\@gobble}%
\providecommand \bibinfo  [0]{\@secondoftwo}%
\providecommand \bibfield  [0]{\@secondoftwo}%
\providecommand \translation [1]{[#1]}%
\providecommand \BibitemOpen [0]{}%
\providecommand \bibitemStop [0]{}%
\providecommand \bibitemNoStop [0]{.\EOS\space}%
\providecommand \EOS [0]{\spacefactor3000\relax}%
\providecommand \BibitemShut  [1]{\csname bibitem#1\endcsname}%
\let\auto@bib@innerbib\@empty
\bibitem [{\citenamefont {Gu}\ \emph {et~al.}(2011)\citenamefont {Gu},
  \citenamefont {Fertig}, \citenamefont {Arovas},\ and\ \citenamefont
  {Auerbach}}]{Gu}%
  \BibitemOpen
  \bibfield  {author} {\bibinfo {author} {\bibfnamefont {Z.}~\bibnamefont
  {Gu}}, \bibinfo {author} {\bibfnamefont {H.~A.}\ \bibnamefont {Fertig}},
  \bibinfo {author} {\bibfnamefont {D.~P.}\ \bibnamefont {Arovas}},\ and\
  \bibinfo {author} {\bibfnamefont {A.}~\bibnamefont {Auerbach}},\ }\bibfield
  {title} {\bibinfo {title} {Floquet spectrum and transport through an
  irradiated graphene ribbon},\ }\href
  {https://doi.org/10.1103/PhysRevLett.107.216601} {\bibfield  {journal}
  {\bibinfo  {journal} {Phys. Rev. Lett.}\ }\textbf {\bibinfo {volume} {107}},\
  \bibinfo {pages} {216601} (\bibinfo {year} {2011})}\BibitemShut {NoStop}%
\bibitem [{\citenamefont {Lindner}\ \emph {et~al.}(2011)\citenamefont
  {Lindner}, \citenamefont {Refael},\ and\ \citenamefont
  {Galitski}}]{Lindner1}%
  \BibitemOpen
  \bibfield  {author} {\bibinfo {author} {\bibfnamefont {N.}~\bibnamefont
  {Lindner}}, \bibinfo {author} {\bibfnamefont {G.}~\bibnamefont {Refael}},\
  and\ \bibinfo {author} {\bibfnamefont {V.}~\bibnamefont {Galitski}},\
  }\bibfield  {title} {\bibinfo {title} {Floquet topological insulator in
  semiconductor quantum wells},\ }\href {https://doi.org/10.1038/nphys1926}
  {\bibfield  {journal} {\bibinfo  {journal} {Nature Phys.}\ }\textbf {\bibinfo
  {volume} {7}},\ \bibinfo {pages} {490–495} (\bibinfo {year}
  {2011})}\BibitemShut {NoStop}%
\bibitem [{\citenamefont {Dehghani}\ \emph {et~al.}(2014)\citenamefont
  {Dehghani}, \citenamefont {Oka},\ and\ \citenamefont {Mitra}}]{Dehghani1}%
  \BibitemOpen
  \bibfield  {author} {\bibinfo {author} {\bibfnamefont {H.}~\bibnamefont
  {Dehghani}}, \bibinfo {author} {\bibfnamefont {T.}~\bibnamefont {Oka}},\ and\
  \bibinfo {author} {\bibfnamefont {A.}~\bibnamefont {Mitra}},\ }\bibfield
  {title} {\bibinfo {title} {Dissipative floquet topological systems},\ }\href
  {https://doi.org/10.1103/PhysRevB.90.195429} {\bibfield  {journal} {\bibinfo
  {journal} {Phys. Rev. B}\ }\textbf {\bibinfo {volume} {90}},\ \bibinfo
  {pages} {195429} (\bibinfo {year} {2014})}\BibitemShut {NoStop}%
\bibitem [{\citenamefont {Dehghani}\ \emph {et~al.}(2015)\citenamefont
  {Dehghani}, \citenamefont {Oka},\ and\ \citenamefont {Mitra}}]{Dehghani2}%
  \BibitemOpen
  \bibfield  {author} {\bibinfo {author} {\bibfnamefont {H.}~\bibnamefont
  {Dehghani}}, \bibinfo {author} {\bibfnamefont {T.}~\bibnamefont {Oka}},\ and\
  \bibinfo {author} {\bibfnamefont {A.}~\bibnamefont {Mitra}},\ }\bibfield
  {title} {\bibinfo {title} {Out-of-equilibrium electrons and the hall
  conductance of a floquet topological insulator},\ }\href
  {https://doi.org/10.1103/PhysRevB.91.155422} {\bibfield  {journal} {\bibinfo
  {journal} {Phys. Rev. B}\ }\textbf {\bibinfo {volume} {91}},\ \bibinfo
  {pages} {155422} (\bibinfo {year} {2015})}\BibitemShut {NoStop}%
\bibitem [{\citenamefont {Farrell}\ and\ \citenamefont
  {Pereg-Barnea}(2016{\natexlab{a}})}]{Farrell1}%
  \BibitemOpen
  \bibfield  {author} {\bibinfo {author} {\bibfnamefont {A.}~\bibnamefont
  {Farrell}}\ and\ \bibinfo {author} {\bibfnamefont {T.}~\bibnamefont
  {Pereg-Barnea}},\ }\bibfield  {title} {\bibinfo {title} {Edge-state transport
  in floquet topological insulators},\ }\href
  {https://doi.org/10.1103/PhysRevB.93.045121} {\bibfield  {journal} {\bibinfo
  {journal} {Phys. Rev. B}\ }\textbf {\bibinfo {volume} {93}},\ \bibinfo
  {pages} {045121} (\bibinfo {year} {2016}{\natexlab{a}})}\BibitemShut
  {NoStop}%
\bibitem [{\citenamefont {Farrell}\ and\ \citenamefont
  {Pereg-Barnea}(2015)}]{Farrell2}%
  \BibitemOpen
  \bibfield  {author} {\bibinfo {author} {\bibfnamefont {A.}~\bibnamefont
  {Farrell}}\ and\ \bibinfo {author} {\bibfnamefont {T.}~\bibnamefont
  {Pereg-Barnea}},\ }\bibfield  {title} {\bibinfo {title} {Photon-inhibited
  topological transport in quantum well heterostructures},\ }\href
  {https://doi.org/10.1103/PhysRevLett.115.106403} {\bibfield  {journal}
  {\bibinfo  {journal} {Phys. Rev. Lett.}\ }\textbf {\bibinfo {volume} {115}},\
  \bibinfo {pages} {106403} (\bibinfo {year} {2015})}\BibitemShut {NoStop}%
\bibitem [{\citenamefont {Kundu}\ and\ \citenamefont {Seradjeh}(2013)}]{Kundu}%
  \BibitemOpen
  \bibfield  {author} {\bibinfo {author} {\bibfnamefont {A.}~\bibnamefont
  {Kundu}}\ and\ \bibinfo {author} {\bibfnamefont {B.}~\bibnamefont
  {Seradjeh}},\ }\bibfield  {title} {\bibinfo {title} {Transport signatures of
  floquet majorana fermions in driven topological superconductors},\ }\href
  {https://doi.org/10.1103/PhysRevLett.111.136402} {\bibfield  {journal}
  {\bibinfo  {journal} {Phys. Rev. Lett.}\ }\textbf {\bibinfo {volume} {111}},\
  \bibinfo {pages} {136402} (\bibinfo {year} {2013})}\BibitemShut {NoStop}%
\bibitem [{\citenamefont {Wang}\ \emph {et~al.}(2013)\citenamefont {Wang},
  \citenamefont {Steinberg}, \citenamefont {Jarillo-Herrero},\ and\
  \citenamefont {Gedik}}]{Wang_2013}%
  \BibitemOpen
  \bibfield  {author} {\bibinfo {author} {\bibfnamefont {Y.~H.}\ \bibnamefont
  {Wang}}, \bibinfo {author} {\bibfnamefont {H.}~\bibnamefont {Steinberg}},
  \bibinfo {author} {\bibfnamefont {P.}~\bibnamefont {Jarillo-Herrero}},\ and\
  \bibinfo {author} {\bibfnamefont {N.}~\bibnamefont {Gedik}},\ }\bibfield
  {title} {\bibinfo {title} {Observation of floquet-bloch states on the surface
  of a topological insulator},\ }\href
  {https://doi.org/10.1126/science.1239834} {\bibfield  {journal} {\bibinfo
  {journal} {Science}\ }\textbf {\bibinfo {volume} {342}},\ \bibinfo {pages}
  {453} (\bibinfo {year} {2013})},\ \Eprint
  {https://arxiv.org/abs/https://www.science.org/doi/pdf/10.1126/science.1239834}
  {https://www.science.org/doi/pdf/10.1126/science.1239834} \BibitemShut
  {NoStop}%
\bibitem [{\citenamefont {Park}\ \emph {et~al.}(2022)\citenamefont {Park},
  \citenamefont {Lee}, \citenamefont {Jang}, \citenamefont {Choi},
  \citenamefont {Park}, \citenamefont {Jung}, \citenamefont {Watanabe},
  \citenamefont {Taniguchi}, \citenamefont {Cho},\ and\ \citenamefont
  {Lee}}]{Lee22}%
  \BibitemOpen
  \bibfield  {author} {\bibinfo {author} {\bibfnamefont {S.}~\bibnamefont
  {Park}}, \bibinfo {author} {\bibfnamefont {W.}~\bibnamefont {Lee}}, \bibinfo
  {author} {\bibfnamefont {S.}~\bibnamefont {Jang}}, \bibinfo {author}
  {\bibfnamefont {Y.-B.}\ \bibnamefont {Choi}}, \bibinfo {author}
  {\bibfnamefont {J.}~\bibnamefont {Park}}, \bibinfo {author} {\bibfnamefont
  {W.}~\bibnamefont {Jung}}, \bibinfo {author} {\bibfnamefont {K.}~\bibnamefont
  {Watanabe}}, \bibinfo {author} {\bibfnamefont {T.}~\bibnamefont {Taniguchi}},
  \bibinfo {author} {\bibfnamefont {G.~Y.}\ \bibnamefont {Cho}},\ and\ \bibinfo
  {author} {\bibfnamefont {G.-H.}\ \bibnamefont {Lee}},\ }\bibfield  {title}
  {\bibinfo {title} {Steady floquet–andreev states in graphene josephson
  junctions},\ }\href@noop {} {\bibfield  {journal} {\bibinfo  {journal}
  {Nature}\ }\textbf {\bibinfo {volume} {603}},\ \bibinfo {pages} {421}
  (\bibinfo {year} {2022})}\BibitemShut {NoStop}%
\bibitem [{\citenamefont {McIver}\ \emph {et~al.}(2020)\citenamefont {McIver},
  \citenamefont {Schulte}, \citenamefont {Stein},\ and\ \citenamefont {{\it et
  al.}}}]{McIver_2020}%
  \BibitemOpen
  \bibfield  {author} {\bibinfo {author} {\bibfnamefont {J.}~\bibnamefont
  {McIver}}, \bibinfo {author} {\bibfnamefont {B.}~\bibnamefont {Schulte}},
  \bibinfo {author} {\bibfnamefont {F.}~\bibnamefont {Stein}},\ and\ \bibinfo
  {author} {\bibnamefont {{\it et al.}}},\ }\bibfield  {title} {\bibinfo
  {title} {Light-induced anomalous hall effect in graphene},\ }\href
  {https://doi.org/10.1038/s41567-019-0698-y} {\bibfield  {journal} {\bibinfo
  {journal} {Nat. Phys.}\ }\textbf {\bibinfo {volume} {16}},\ \bibinfo {pages}
  {38–41} (\bibinfo {year} {2020})}\BibitemShut {NoStop}%
\bibitem [{\citenamefont {Rechtsman}\ \emph {et~al.}(2013)\citenamefont
  {Rechtsman}, \citenamefont {Zeuner}, \citenamefont {Plotnik},\ and\
  \citenamefont {{\it et al.}}}]{Rechtsman_2013}%
  \BibitemOpen
  \bibfield  {author} {\bibinfo {author} {\bibfnamefont {M.}~\bibnamefont
  {Rechtsman}}, \bibinfo {author} {\bibfnamefont {J.}~\bibnamefont {Zeuner}},
  \bibinfo {author} {\bibfnamefont {Y.}~\bibnamefont {Plotnik}},\ and\ \bibinfo
  {author} {\bibnamefont {{\it et al.}}},\ }\bibfield  {title} {\bibinfo
  {title} {Photonic floquet topological insulators},\ }\href
  {https://doi.org/10.1038/nature12066} {\bibfield  {journal} {\bibinfo
  {journal} {Nature.}\ }\textbf {\bibinfo {volume} {496}},\ \bibinfo {pages}
  {196–200} (\bibinfo {year} {2013})}\BibitemShut {NoStop}%
\bibitem [{\citenamefont {Mukherjee}\ and\ \citenamefont
  {Rechtsman}(2021)}]{Mukherjee_2021}%
  \BibitemOpen
  \bibfield  {author} {\bibinfo {author} {\bibfnamefont {S.}~\bibnamefont
  {Mukherjee}}\ and\ \bibinfo {author} {\bibfnamefont {M.~C.}\ \bibnamefont
  {Rechtsman}},\ }\bibfield  {title} {\bibinfo {title} {Observation of
  unidirectional solitonlike edge states in nonlinear floquet topological
  insulators},\ }\href {https://doi.org/10.1103/PhysRevX.11.041057} {\bibfield
  {journal} {\bibinfo  {journal} {Phys. Rev. X}\ }\textbf {\bibinfo {volume}
  {11}},\ \bibinfo {pages} {041057} (\bibinfo {year} {2021})}\BibitemShut
  {NoStop}%
\bibitem [{\citenamefont {Guglielmon}\ \emph {et~al.}(2018)\citenamefont
  {Guglielmon}, \citenamefont {Huang}, \citenamefont {Chen},\ and\
  \citenamefont {Rechtsman}}]{Guglielmon_2018}%
  \BibitemOpen
  \bibfield  {author} {\bibinfo {author} {\bibfnamefont {J.}~\bibnamefont
  {Guglielmon}}, \bibinfo {author} {\bibfnamefont {S.}~\bibnamefont {Huang}},
  \bibinfo {author} {\bibfnamefont {K.~P.}\ \bibnamefont {Chen}},\ and\
  \bibinfo {author} {\bibfnamefont {M.~C.}\ \bibnamefont {Rechtsman}},\
  }\bibfield  {title} {\bibinfo {title} {Photonic realization of a transition
  to a strongly driven floquet topological phase},\ }\href
  {https://doi.org/10.1103/PhysRevA.97.031801} {\bibfield  {journal} {\bibinfo
  {journal} {Phys. Rev. A}\ }\textbf {\bibinfo {volume} {97}},\ \bibinfo
  {pages} {031801} (\bibinfo {year} {2018})}\BibitemShut {NoStop}%
\bibitem [{\citenamefont {Alicea}\ \emph {et~al.}(2011)\citenamefont {Alicea},
  \citenamefont {Oreg}, \citenamefont {Refael},\ and\ \citenamefont {{\it et
  al.}}}]{Alicea_2011}%
  \BibitemOpen
  \bibfield  {author} {\bibinfo {author} {\bibfnamefont {J.}~\bibnamefont
  {Alicea}}, \bibinfo {author} {\bibfnamefont {Y.}~\bibnamefont {Oreg}},
  \bibinfo {author} {\bibfnamefont {G.}~\bibnamefont {Refael}},\ and\ \bibinfo
  {author} {\bibnamefont {{\it et al.}}},\ }\bibfield  {title} {\bibinfo
  {title} {Non-abelian statistics and topological quantum information
  processing in 1d wire networks},\ }\href {https://doi.org/10.1038/nphys1915}
  {\bibfield  {journal} {\bibinfo  {journal} {Nature Phys.}\ }\textbf {\bibinfo
  {volume} {7}},\ \bibinfo {pages} {412–417} (\bibinfo {year}
  {2011})}\BibitemShut {NoStop}%
\bibitem [{\citenamefont {Plugge}\ \emph {et~al.}(2017)\citenamefont {Plugge},
  \citenamefont {Rasmussen}, \citenamefont {Egger},\ and\ \citenamefont
  {Flensberg}}]{Plugge_2010}%
  \BibitemOpen
  \bibfield  {author} {\bibinfo {author} {\bibfnamefont {S.}~\bibnamefont
  {Plugge}}, \bibinfo {author} {\bibfnamefont {A.}~\bibnamefont {Rasmussen}},
  \bibinfo {author} {\bibfnamefont {R.}~\bibnamefont {Egger}},\ and\ \bibinfo
  {author} {\bibfnamefont {K.}~\bibnamefont {Flensberg}},\ }\bibfield  {title}
  {\bibinfo {title} {Majorana box qubits},\ }\href
  {https://doi.org/10.1088/1367-2630/aa54e1} {\bibfield  {journal} {\bibinfo
  {journal} {New J. Phys.}\ }\textbf {\bibinfo {volume} {19}},\ \bibinfo
  {pages} {012001} (\bibinfo {year} {2017})}\BibitemShut {NoStop}%
\bibitem [{\citenamefont {Kitaev}(2001)}]{Kitaev_2001}%
  \BibitemOpen
  \bibfield  {author} {\bibinfo {author} {\bibfnamefont {A.~Y.}\ \bibnamefont
  {Kitaev}},\ }\bibfield  {title} {\bibinfo {title} {Unpaired majorana fermions
  in quantum wires},\ }\href {https://doi.org/10.1070/1063-7869/44/10S/S29}
  {\bibfield  {journal} {\bibinfo  {journal} {Phys.-Usp.}\ }\textbf {\bibinfo
  {volume} {44}},\ \bibinfo {pages} {131} (\bibinfo {year} {2001})}\BibitemShut
  {NoStop}%
\bibitem [{\citenamefont {Lutchyn}\ \emph {et~al.}(2010)\citenamefont
  {Lutchyn}, \citenamefont {Sau},\ and\ \citenamefont
  {Das~Sarma}}]{Lutchyn_2010}%
  \BibitemOpen
  \bibfield  {author} {\bibinfo {author} {\bibfnamefont {R.~M.}\ \bibnamefont
  {Lutchyn}}, \bibinfo {author} {\bibfnamefont {J.~D.}\ \bibnamefont {Sau}},\
  and\ \bibinfo {author} {\bibfnamefont {S.}~\bibnamefont {Das~Sarma}},\
  }\bibfield  {title} {\bibinfo {title} {Majorana fermions and a topological
  phase transition in semiconductor-superconductor heterostructures},\ }\href
  {https://doi.org/10.1103/PhysRevLett.105.077001} {\bibfield  {journal}
  {\bibinfo  {journal} {Phys. Rev. Lett.}\ }\textbf {\bibinfo {volume} {105}},\
  \bibinfo {pages} {077001} (\bibinfo {year} {2010})}\BibitemShut {NoStop}%
\bibitem [{\citenamefont {Oreg}\ \emph {et~al.}(2010)\citenamefont {Oreg},
  \citenamefont {Refael},\ and\ \citenamefont {von Oppen}}]{Oreg_2010}%
  \BibitemOpen
  \bibfield  {author} {\bibinfo {author} {\bibfnamefont {Y.}~\bibnamefont
  {Oreg}}, \bibinfo {author} {\bibfnamefont {G.}~\bibnamefont {Refael}},\ and\
  \bibinfo {author} {\bibfnamefont {F.}~\bibnamefont {von Oppen}},\ }\bibfield
  {title} {\bibinfo {title} {Helical liquids and majorana bound states in
  quantum wires},\ }\href {https://doi.org/10.1103/PhysRevLett.105.177002}
  {\bibfield  {journal} {\bibinfo  {journal} {Phys. Rev. Lett.}\ }\textbf
  {\bibinfo {volume} {105}},\ \bibinfo {pages} {177002} (\bibinfo {year}
  {2010})}\BibitemShut {NoStop}%
\bibitem [{\citenamefont {Ryu}\ \emph {et~al.}(2010)\citenamefont {Ryu},
  \citenamefont {Schnyder}, \citenamefont {Furusaki},\ and\ \citenamefont
  {Ludwig}}]{Ryu10}%
  \BibitemOpen
  \bibfield  {author} {\bibinfo {author} {\bibfnamefont {S.}~\bibnamefont
  {Ryu}}, \bibinfo {author} {\bibfnamefont {A.~P.}\ \bibnamefont {Schnyder}},
  \bibinfo {author} {\bibfnamefont {A.}~\bibnamefont {Furusaki}},\ and\
  \bibinfo {author} {\bibfnamefont {A.~W.~W.}\ \bibnamefont {Ludwig}},\
  }\bibfield  {title} {\bibinfo {title} {Topological insulators and
  superconductors: tenfold way and dimensional hierarchy},\ }\href
  {http://stacks.iop.org/1367-2630/12/i=6/a=065010} {\bibfield  {journal}
  {\bibinfo  {journal} {New Journal of Physics}\ }\textbf {\bibinfo {volume}
  {12}},\ \bibinfo {pages} {065010} (\bibinfo {year} {2010})}\BibitemShut
  {NoStop}%
\bibitem [{\citenamefont {Kitaev}\ and\ \citenamefont
  {Preskill}(2006)}]{Preskill06}%
  \BibitemOpen
  \bibfield  {author} {\bibinfo {author} {\bibfnamefont {A.}~\bibnamefont
  {Kitaev}}\ and\ \bibinfo {author} {\bibfnamefont {J.}~\bibnamefont
  {Preskill}},\ }\bibfield  {title} {\bibinfo {title} {Topological entanglement
  entropy},\ }\href {https://doi.org/10.1103/PhysRevLett.96.110404} {\bibfield
  {journal} {\bibinfo  {journal} {Phys. Rev. Lett.}\ }\textbf {\bibinfo
  {volume} {96}},\ \bibinfo {pages} {110404} (\bibinfo {year}
  {2006})}\BibitemShut {NoStop}%
\bibitem [{\citenamefont {Levin}\ and\ \citenamefont {Wen}(2006)}]{Levin2006}%
  \BibitemOpen
  \bibfield  {author} {\bibinfo {author} {\bibfnamefont {M.}~\bibnamefont
  {Levin}}\ and\ \bibinfo {author} {\bibfnamefont {X.-G.}\ \bibnamefont
  {Wen}},\ }\bibfield  {title} {\bibinfo {title} {Detecting topological order
  in a ground state wave function},\ }\href
  {https://doi.org/10.1103/PhysRevLett.96.110405} {\bibfield  {journal}
  {\bibinfo  {journal} {Phys. Rev. Lett.}\ }\textbf {\bibinfo {volume} {96}},\
  \bibinfo {pages} {110405} (\bibinfo {year} {2006})}\BibitemShut {NoStop}%
\bibitem [{\citenamefont {Borchmann}\ \emph {et~al.}(2014)\citenamefont
  {Borchmann}, \citenamefont {Farrell}, \citenamefont {Matsuura},\ and\
  \citenamefont {Pereg-Barnea}}]{Borchmann_2014}%
  \BibitemOpen
  \bibfield  {author} {\bibinfo {author} {\bibfnamefont {J.}~\bibnamefont
  {Borchmann}}, \bibinfo {author} {\bibfnamefont {A.}~\bibnamefont {Farrell}},
  \bibinfo {author} {\bibfnamefont {S.}~\bibnamefont {Matsuura}},\ and\
  \bibinfo {author} {\bibfnamefont {T.}~\bibnamefont {Pereg-Barnea}},\
  }\bibfield  {title} {\bibinfo {title} {Entanglement spectrum as a probe for
  the topology of a spin-orbit-coupled superconductor},\ }\href
  {https://doi.org/10.1103/PhysRevB.90.235150} {\bibfield  {journal} {\bibinfo
  {journal} {Phys. Rev. B}\ }\textbf {\bibinfo {volume} {90}},\ \bibinfo
  {pages} {235150} (\bibinfo {year} {2014})}\BibitemShut {NoStop}%
\bibitem [{\citenamefont {Borchmann}\ \emph {et~al.}(2016)\citenamefont
  {Borchmann}, \citenamefont {Farrell},\ and\ \citenamefont
  {Pereg-Barnea}}]{Borchmann_2016}%
  \BibitemOpen
  \bibfield  {author} {\bibinfo {author} {\bibfnamefont {J.}~\bibnamefont
  {Borchmann}}, \bibinfo {author} {\bibfnamefont {A.}~\bibnamefont {Farrell}},\
  and\ \bibinfo {author} {\bibfnamefont {T.}~\bibnamefont {Pereg-Barnea}},\
  }\bibfield  {title} {\bibinfo {title} {Anderson topological superconductor},\
  }\href {https://doi.org/10.1103/PhysRevB.93.125133} {\bibfield  {journal}
  {\bibinfo  {journal} {Phys. Rev. B}\ }\textbf {\bibinfo {volume} {93}},\
  \bibinfo {pages} {125133} (\bibinfo {year} {2016})}\BibitemShut {NoStop}%
\bibitem [{\citenamefont {Borchmann}\ and\ \citenamefont
  {Pereg-Barnea}(2017)}]{Borchmann_2017}%
  \BibitemOpen
  \bibfield  {author} {\bibinfo {author} {\bibfnamefont {J.}~\bibnamefont
  {Borchmann}}\ and\ \bibinfo {author} {\bibfnamefont {T.}~\bibnamefont
  {Pereg-Barnea}},\ }\bibfield  {title} {\bibinfo {title} {Analytic expression
  for the entanglement entropy of a two-dimensional topological
  superconductor},\ }\href {https://doi.org/10.1103/PhysRevB.95.075152}
  {\bibfield  {journal} {\bibinfo  {journal} {Phys. Rev. B}\ }\textbf {\bibinfo
  {volume} {95}},\ \bibinfo {pages} {075152} (\bibinfo {year}
  {2017})}\BibitemShut {NoStop}%
\bibitem [{\citenamefont {Zeng}\ and\ \citenamefont {Wen}(2015)}]{Zeng2015}%
  \BibitemOpen
  \bibfield  {author} {\bibinfo {author} {\bibfnamefont {B.}~\bibnamefont
  {Zeng}}\ and\ \bibinfo {author} {\bibfnamefont {X.-G.}\ \bibnamefont {Wen}},\
  }\bibfield  {title} {\bibinfo {title} {Gapped quantum liquids and topological
  order, stochastic local transformations and emergence of unitarity},\ }\href
  {https://doi.org/10.1103/PhysRevB.91.125121} {\bibfield  {journal} {\bibinfo
  {journal} {Phys. Rev. B}\ }\textbf {\bibinfo {volume} {91}},\ \bibinfo
  {pages} {125121} (\bibinfo {year} {2015})}\BibitemShut {NoStop}%
\bibitem [{\citenamefont {Zeng}\ and\ \citenamefont {Zhou}(2016)}]{Zeng_2016}%
  \BibitemOpen
  \bibfield  {author} {\bibinfo {author} {\bibfnamefont {B.}~\bibnamefont
  {Zeng}}\ and\ \bibinfo {author} {\bibfnamefont {D.~L.}\ \bibnamefont
  {Zhou}},\ }\bibfield  {title} {\bibinfo {title} {Topological and
  error-correcting properties for symmetry-protected topological order},\
  }\href {https://doi.org/10.1209/0295-5075/113/56001} {\bibfield  {journal}
  {\bibinfo  {journal} {Europhysics Letters}\ }\textbf {\bibinfo {volume}
  {113}},\ \bibinfo {pages} {56001} (\bibinfo {year} {2016})}\BibitemShut
  {NoStop}%
\bibitem [{\citenamefont {Fromholz}\ \emph {et~al.}(2020)\citenamefont
  {Fromholz}, \citenamefont {Magnifico}, \citenamefont {Vitale}, \citenamefont
  {Mendes-Santos},\ and\ \citenamefont {Dalmonte}}]{Fromholz_2020}%
  \BibitemOpen
  \bibfield  {author} {\bibinfo {author} {\bibfnamefont {P.}~\bibnamefont
  {Fromholz}}, \bibinfo {author} {\bibfnamefont {G.}~\bibnamefont {Magnifico}},
  \bibinfo {author} {\bibfnamefont {V.}~\bibnamefont {Vitale}}, \bibinfo
  {author} {\bibfnamefont {T.}~\bibnamefont {Mendes-Santos}},\ and\ \bibinfo
  {author} {\bibfnamefont {M.}~\bibnamefont {Dalmonte}},\ }\bibfield  {title}
  {\bibinfo {title} {Entanglement topological invariants for one-dimensional
  topological superconductors},\ }\href
  {https://doi.org/10.1103/PhysRevB.101.085136} {\bibfield  {journal} {\bibinfo
   {journal} {Phys. Rev. B}\ }\textbf {\bibinfo {volume} {101}},\ \bibinfo
  {pages} {085136} (\bibinfo {year} {2020})}\BibitemShut {NoStop}%
\bibitem [{\citenamefont {Khemani}\ \emph {et~al.}(2016)\citenamefont
  {Khemani}, \citenamefont {Lazarides}, \citenamefont {Moessner},\ and\
  \citenamefont {Sondhi}}]{Khemani_2016}%
  \BibitemOpen
  \bibfield  {author} {\bibinfo {author} {\bibfnamefont {V.}~\bibnamefont
  {Khemani}}, \bibinfo {author} {\bibfnamefont {A.}~\bibnamefont {Lazarides}},
  \bibinfo {author} {\bibfnamefont {R.}~\bibnamefont {Moessner}},\ and\
  \bibinfo {author} {\bibfnamefont {S.~L.}\ \bibnamefont {Sondhi}},\ }\bibfield
   {title} {\bibinfo {title} {Phase structure of driven quantum systems},\
  }\href {https://doi.org/10.1103/PhysRevLett.116.250401} {\bibfield  {journal}
  {\bibinfo  {journal} {Phys. Rev. Lett.}\ }\textbf {\bibinfo {volume} {116}},\
  \bibinfo {pages} {250401} (\bibinfo {year} {2016})}\BibitemShut {NoStop}%
\bibitem [{\citenamefont {Jiang}\ \emph {et~al.}(2011)\citenamefont {Jiang},
  \citenamefont {Kitagawa}, \citenamefont {Alicea}, \citenamefont {Akhmerov},
  \citenamefont {Pekker}, \citenamefont {Refael}, \citenamefont {Cirac},
  \citenamefont {Demler}, \citenamefont {Lukin},\ and\ \citenamefont
  {Zoller}}]{Jiang_2011}%
  \BibitemOpen
  \bibfield  {author} {\bibinfo {author} {\bibfnamefont {L.}~\bibnamefont
  {Jiang}}, \bibinfo {author} {\bibfnamefont {T.}~\bibnamefont {Kitagawa}},
  \bibinfo {author} {\bibfnamefont {J.}~\bibnamefont {Alicea}}, \bibinfo
  {author} {\bibfnamefont {A.~R.}\ \bibnamefont {Akhmerov}}, \bibinfo {author}
  {\bibfnamefont {D.}~\bibnamefont {Pekker}}, \bibinfo {author} {\bibfnamefont
  {G.}~\bibnamefont {Refael}}, \bibinfo {author} {\bibfnamefont {J.~I.}\
  \bibnamefont {Cirac}}, \bibinfo {author} {\bibfnamefont {E.}~\bibnamefont
  {Demler}}, \bibinfo {author} {\bibfnamefont {M.~D.}\ \bibnamefont {Lukin}},\
  and\ \bibinfo {author} {\bibfnamefont {P.}~\bibnamefont {Zoller}},\
  }\bibfield  {title} {\bibinfo {title} {Majorana fermions in equilibrium and
  in driven cold-atom quantum wires},\ }\href
  {https://doi.org/10.1103/PhysRevLett.106.220402} {\bibfield  {journal}
  {\bibinfo  {journal} {Phys. Rev. Lett.}\ }\textbf {\bibinfo {volume} {106}},\
  \bibinfo {pages} {220402} (\bibinfo {year} {2011})}\BibitemShut {NoStop}%
\bibitem [{\citenamefont {Bauer}\ \emph {et~al.}(2019)\citenamefont {Bauer},
  \citenamefont {Pereg-Barnea}, \citenamefont {Karzig}, \citenamefont {Rieder},
  \citenamefont {Refael}, \citenamefont {Berg},\ and\ \citenamefont
  {Oreg}}]{Bauer_2019}%
  \BibitemOpen
  \bibfield  {author} {\bibinfo {author} {\bibfnamefont {B.}~\bibnamefont
  {Bauer}}, \bibinfo {author} {\bibfnamefont {T.}~\bibnamefont {Pereg-Barnea}},
  \bibinfo {author} {\bibfnamefont {T.}~\bibnamefont {Karzig}}, \bibinfo
  {author} {\bibfnamefont {M.-T.}\ \bibnamefont {Rieder}}, \bibinfo {author}
  {\bibfnamefont {G.}~\bibnamefont {Refael}}, \bibinfo {author} {\bibfnamefont
  {E.}~\bibnamefont {Berg}},\ and\ \bibinfo {author} {\bibfnamefont
  {Y.}~\bibnamefont {Oreg}},\ }\bibfield  {title} {\bibinfo {title}
  {Topologically protected braiding in a single wire using floquet majorana
  modes},\ }\href {https://doi.org/10.1103/PhysRevB.100.041102} {\bibfield
  {journal} {\bibinfo  {journal} {Phys. Rev. B}\ }\textbf {\bibinfo {volume}
  {100}},\ \bibinfo {pages} {041102} (\bibinfo {year} {2019})}\BibitemShut
  {NoStop}%
\bibitem [{\citenamefont {Asb\'oth}\ and\ \citenamefont
  {Obuse}(2013)}]{Obuse13}%
  \BibitemOpen
  \bibfield  {author} {\bibinfo {author} {\bibfnamefont {J.~K.}\ \bibnamefont
  {Asb\'oth}}\ and\ \bibinfo {author} {\bibfnamefont {H.}~\bibnamefont
  {Obuse}},\ }\bibfield  {title} {\bibinfo {title} {Bulk-boundary
  correspondence for chiral symmetric quantum walks},\ }\href
  {https://doi.org/10.1103/PhysRevB.88.121406} {\bibfield  {journal} {\bibinfo
  {journal} {Phys. Rev. B}\ }\textbf {\bibinfo {volume} {88}},\ \bibinfo
  {pages} {121406} (\bibinfo {year} {2013})}\BibitemShut {NoStop}%
\bibitem [{\citenamefont {Asb\'oth}\ \emph {et~al.}(2014)\citenamefont
  {Asb\'oth}, \citenamefont {Tarasinski},\ and\ \citenamefont
  {Delplace}}]{Delplace14}%
  \BibitemOpen
  \bibfield  {author} {\bibinfo {author} {\bibfnamefont {J.~K.}\ \bibnamefont
  {Asb\'oth}}, \bibinfo {author} {\bibfnamefont {B.}~\bibnamefont
  {Tarasinski}},\ and\ \bibinfo {author} {\bibfnamefont {P.}~\bibnamefont
  {Delplace}},\ }\bibfield  {title} {\bibinfo {title} {Chiral symmetry and
  bulk-boundary correspondence in periodically driven one-dimensional
  systems},\ }\href {https://doi.org/10.1103/PhysRevB.90.125143} {\bibfield
  {journal} {\bibinfo  {journal} {Phys. Rev. B}\ }\textbf {\bibinfo {volume}
  {90}},\ \bibinfo {pages} {125143} (\bibinfo {year} {2014})}\BibitemShut
  {NoStop}%
\bibitem [{\citenamefont {Yates}\ and\ \citenamefont {Mitra}(2017)}]{Yates17}%
  \BibitemOpen
  \bibfield  {author} {\bibinfo {author} {\bibfnamefont {D.~J.}\ \bibnamefont
  {Yates}}\ and\ \bibinfo {author} {\bibfnamefont {A.}~\bibnamefont {Mitra}},\
  }\bibfield  {title} {\bibinfo {title} {Entanglement properties of the
  time-periodic kitaev chain},\ }\href
  {https://doi.org/10.1103/PhysRevB.96.115108} {\bibfield  {journal} {\bibinfo
  {journal} {Phys. Rev. B}\ }\textbf {\bibinfo {volume} {96}},\ \bibinfo
  {pages} {115108} (\bibinfo {year} {2017})}\BibitemShut {NoStop}%
\bibitem [{\citenamefont {Yates}\ \emph {et~al.}(2018)\citenamefont {Yates},
  \citenamefont {Lemonik},\ and\ \citenamefont {Mitra}}]{Yates18}%
  \BibitemOpen
  \bibfield  {author} {\bibinfo {author} {\bibfnamefont {D.}~\bibnamefont
  {Yates}}, \bibinfo {author} {\bibfnamefont {Y.}~\bibnamefont {Lemonik}},\
  and\ \bibinfo {author} {\bibfnamefont {A.}~\bibnamefont {Mitra}},\ }\bibfield
   {title} {\bibinfo {title} {Central charge of periodically driven critical
  kitaev chains},\ }\href {https://doi.org/10.1103/PhysRevLett.121.076802}
  {\bibfield  {journal} {\bibinfo  {journal} {Phys. Rev. Lett.}\ }\textbf
  {\bibinfo {volume} {121}},\ \bibinfo {pages} {076802} (\bibinfo {year}
  {2018})}\BibitemShut {NoStop}%
\bibitem [{\citenamefont {Motrunich}\ \emph {et~al.}(2001)\citenamefont
  {Motrunich}, \citenamefont {Damle},\ and\ \citenamefont
  {Huse}}]{Motrunich2001}%
  \BibitemOpen
  \bibfield  {author} {\bibinfo {author} {\bibfnamefont {O.}~\bibnamefont
  {Motrunich}}, \bibinfo {author} {\bibfnamefont {K.}~\bibnamefont {Damle}},\
  and\ \bibinfo {author} {\bibfnamefont {D.~A.}\ \bibnamefont {Huse}},\
  }\bibfield  {title} {\bibinfo {title} {Griffiths effects and quantum critical
  points in dirty superconductors without spin-rotation invariance:
  One-dimensional examples},\ }\href
  {https://doi.org/10.1103/PhysRevB.63.224204} {\bibfield  {journal} {\bibinfo
  {journal} {Phys. Rev. B}\ }\textbf {\bibinfo {volume} {63}},\ \bibinfo
  {pages} {224204} (\bibinfo {year} {2001})}\BibitemShut {NoStop}%
\bibitem [{\citenamefont {Brouwer}\ \emph
  {et~al.}(2011{\natexlab{a}})\citenamefont {Brouwer}, \citenamefont
  {Duckheim}, \citenamefont {Romito},\ and\ \citenamefont {von
  Oppen}}]{Brouwer2011}%
  \BibitemOpen
  \bibfield  {author} {\bibinfo {author} {\bibfnamefont {P.~W.}\ \bibnamefont
  {Brouwer}}, \bibinfo {author} {\bibfnamefont {M.}~\bibnamefont {Duckheim}},
  \bibinfo {author} {\bibfnamefont {A.}~\bibnamefont {Romito}},\ and\ \bibinfo
  {author} {\bibfnamefont {F.}~\bibnamefont {von Oppen}},\ }\bibfield  {title}
  {\bibinfo {title} {Probability distribution of majorana end-state energies in
  disordered wires},\ }\href {https://doi.org/10.1103/PhysRevLett.107.196804}
  {\bibfield  {journal} {\bibinfo  {journal} {Phys. Rev. Lett.}\ }\textbf
  {\bibinfo {volume} {107}},\ \bibinfo {pages} {196804} (\bibinfo {year}
  {2011}{\natexlab{a}})}\BibitemShut {NoStop}%
\bibitem [{\citenamefont {Brouwer}\ \emph
  {et~al.}(2011{\natexlab{b}})\citenamefont {Brouwer}, \citenamefont
  {Duckheim}, \citenamefont {Romito},\ and\ \citenamefont {von
  Oppen}}]{Brouwer2011a}%
  \BibitemOpen
  \bibfield  {author} {\bibinfo {author} {\bibfnamefont {P.~W.}\ \bibnamefont
  {Brouwer}}, \bibinfo {author} {\bibfnamefont {M.}~\bibnamefont {Duckheim}},
  \bibinfo {author} {\bibfnamefont {A.}~\bibnamefont {Romito}},\ and\ \bibinfo
  {author} {\bibfnamefont {F.}~\bibnamefont {von Oppen}},\ }\bibfield  {title}
  {\bibinfo {title} {Topological superconducting phases in disordered quantum
  wires with strong spin-orbit coupling},\ }\href
  {https://doi.org/10.1103/PhysRevB.84.144526} {\bibfield  {journal} {\bibinfo
  {journal} {Phys. Rev. B}\ }\textbf {\bibinfo {volume} {84}},\ \bibinfo
  {pages} {144526} (\bibinfo {year} {2011}{\natexlab{b}})}\BibitemShut
  {NoStop}%
\bibitem [{\citenamefont {Akhmerov}\ \emph {et~al.}(2011)\citenamefont
  {Akhmerov}, \citenamefont {Dahlhaus}, \citenamefont {Hassler}, \citenamefont
  {Wimmer},\ and\ \citenamefont {Beenakker}}]{Akhmerov2011}%
  \BibitemOpen
  \bibfield  {author} {\bibinfo {author} {\bibfnamefont {A.~R.}\ \bibnamefont
  {Akhmerov}}, \bibinfo {author} {\bibfnamefont {J.~P.}\ \bibnamefont
  {Dahlhaus}}, \bibinfo {author} {\bibfnamefont {F.}~\bibnamefont {Hassler}},
  \bibinfo {author} {\bibfnamefont {M.}~\bibnamefont {Wimmer}},\ and\ \bibinfo
  {author} {\bibfnamefont {C.~W.~J.}\ \bibnamefont {Beenakker}},\ }\bibfield
  {title} {\bibinfo {title} {Quantized conductance at the majorana phase
  transition in a disordered superconducting wire},\ }\href
  {https://doi.org/10.1103/PhysRevLett.106.057001} {\bibfield  {journal}
  {\bibinfo  {journal} {Phys. Rev. Lett.}\ }\textbf {\bibinfo {volume} {106}},\
  \bibinfo {pages} {057001} (\bibinfo {year} {2011})}\BibitemShut {NoStop}%
\bibitem [{\citenamefont {Rieder}\ \emph {et~al.}(2012)\citenamefont {Rieder},
  \citenamefont {Kells}, \citenamefont {Duckheim}, \citenamefont {Meidan},\
  and\ \citenamefont {Brouwer}}]{Rieder2012}%
  \BibitemOpen
  \bibfield  {author} {\bibinfo {author} {\bibfnamefont {M.-T.}\ \bibnamefont
  {Rieder}}, \bibinfo {author} {\bibfnamefont {G.}~\bibnamefont {Kells}},
  \bibinfo {author} {\bibfnamefont {M.}~\bibnamefont {Duckheim}}, \bibinfo
  {author} {\bibfnamefont {D.}~\bibnamefont {Meidan}},\ and\ \bibinfo {author}
  {\bibfnamefont {P.~W.}\ \bibnamefont {Brouwer}},\ }\bibfield  {title}
  {\bibinfo {title} {Endstates in multichannel spinless $p$-wave
  superconducting wires},\ }\href {https://doi.org/10.1103/PhysRevB.86.125423}
  {\bibfield  {journal} {\bibinfo  {journal} {Phys. Rev. B}\ }\textbf {\bibinfo
  {volume} {86}},\ \bibinfo {pages} {125423} (\bibinfo {year}
  {2012})}\BibitemShut {NoStop}%
\bibitem [{\citenamefont {Rieder}\ \emph {et~al.}(2013)\citenamefont {Rieder},
  \citenamefont {Brouwer},\ and\ \citenamefont {Adagideli}}]{Rieder2013}%
  \BibitemOpen
  \bibfield  {author} {\bibinfo {author} {\bibfnamefont {M.-T.}\ \bibnamefont
  {Rieder}}, \bibinfo {author} {\bibfnamefont {P.~W.}\ \bibnamefont
  {Brouwer}},\ and\ \bibinfo {author} {\bibfnamefont {i.~d. I. m.~c.}\
  \bibnamefont {Adagideli}},\ }\bibfield  {title} {\bibinfo {title} {Reentrant
  topological phase transitions in a disordered spinless superconducting
  wire},\ }\href {https://doi.org/10.1103/PhysRevB.88.060509} {\bibfield
  {journal} {\bibinfo  {journal} {Phys. Rev. B}\ }\textbf {\bibinfo {volume}
  {88}},\ \bibinfo {pages} {060509} (\bibinfo {year} {2013})}\BibitemShut
  {NoStop}%
\bibitem [{\citenamefont {DeGottardi}\ \emph {et~al.}(2013)\citenamefont
  {DeGottardi}, \citenamefont {Sen},\ and\ \citenamefont
  {Vishveshwara}}]{DeGottardi2013}%
  \BibitemOpen
  \bibfield  {author} {\bibinfo {author} {\bibfnamefont {W.}~\bibnamefont
  {DeGottardi}}, \bibinfo {author} {\bibfnamefont {D.}~\bibnamefont {Sen}},\
  and\ \bibinfo {author} {\bibfnamefont {S.}~\bibnamefont {Vishveshwara}},\
  }\bibfield  {title} {\bibinfo {title} {Majorana fermions in superconducting
  1d systems having periodic, quasiperiodic, and disordered potentials},\
  }\href {https://doi.org/10.1103/PhysRevLett.110.146404} {\bibfield  {journal}
  {\bibinfo  {journal} {Phys. Rev. Lett.}\ }\textbf {\bibinfo {volume} {110}},\
  \bibinfo {pages} {146404} (\bibinfo {year} {2013})}\BibitemShut {NoStop}%
\bibitem [{\citenamefont {Pan}\ \emph {et~al.}(2020)\citenamefont {Pan},
  \citenamefont {Cole}, \citenamefont {Sau},\ and\ \citenamefont
  {Das~Sarma}}]{Pan2020zerobiasconductance}%
  \BibitemOpen
  \bibfield  {author} {\bibinfo {author} {\bibfnamefont {H.}~\bibnamefont
  {Pan}}, \bibinfo {author} {\bibfnamefont {W.~S.}\ \bibnamefont {Cole}},
  \bibinfo {author} {\bibfnamefont {J.~D.}\ \bibnamefont {Sau}},\ and\ \bibinfo
  {author} {\bibfnamefont {S.}~\bibnamefont {Das~Sarma}},\ }\bibfield  {title}
  {\bibinfo {title} {Generic quantized zero-bias conductance peaks in
  superconductor-semiconductor hybrid structures},\ }\href
  {https://doi.org/10.1103/PhysRevB.101.024506} {\bibfield  {journal} {\bibinfo
   {journal} {Phys. Rev. B}\ }\textbf {\bibinfo {volume} {101}},\ \bibinfo
  {pages} {024506} (\bibinfo {year} {2020})}\BibitemShut {NoStop}%
\bibitem [{\citenamefont {Pan}\ and\ \citenamefont
  {Das~Sarma}(2021{\natexlab{a}})}]{Pan2021DisorderMZM}%
  \BibitemOpen
  \bibfield  {author} {\bibinfo {author} {\bibfnamefont {H.}~\bibnamefont
  {Pan}}\ and\ \bibinfo {author} {\bibfnamefont {S.}~\bibnamefont
  {Das~Sarma}},\ }\bibfield  {title} {\bibinfo {title} {Disorder effects on
  majorana zero modes: Kitaev chain versus semiconductor nanowire},\ }\href
  {https://doi.org/10.1103/PhysRevB.103.224505} {\bibfield  {journal} {\bibinfo
   {journal} {Phys. Rev. B}\ }\textbf {\bibinfo {volume} {103}},\ \bibinfo
  {pages} {224505} (\bibinfo {year} {2021}{\natexlab{a}})}\BibitemShut
  {NoStop}%
\bibitem [{\citenamefont {Das~Sarma}(2023)}]{das2023search}%
  \BibitemOpen
  \bibfield  {author} {\bibinfo {author} {\bibfnamefont {S.}~\bibnamefont
  {Das~Sarma}},\ }\bibfield  {title} {\bibinfo {title} {In search of
  majorana},\ }\href@noop {} {\bibfield  {journal} {\bibinfo  {journal} {Nature
  Physics}\ }\textbf {\bibinfo {volume} {19}},\ \bibinfo {pages} {165}
  (\bibinfo {year} {2023})}\BibitemShut {NoStop}%
\bibitem [{\citenamefont {Huse}\ \emph {et~al.}(2013)\citenamefont {Huse},
  \citenamefont {Nandkishore}, \citenamefont {Oganesyan}, \citenamefont {Pal},\
  and\ \citenamefont {Sondhi}}]{Huse_2013}%
  \BibitemOpen
  \bibfield  {author} {\bibinfo {author} {\bibfnamefont {D.~A.}\ \bibnamefont
  {Huse}}, \bibinfo {author} {\bibfnamefont {R.}~\bibnamefont {Nandkishore}},
  \bibinfo {author} {\bibfnamefont {V.}~\bibnamefont {Oganesyan}}, \bibinfo
  {author} {\bibfnamefont {A.}~\bibnamefont {Pal}},\ and\ \bibinfo {author}
  {\bibfnamefont {S.~L.}\ \bibnamefont {Sondhi}},\ }\bibfield  {title}
  {\bibinfo {title} {Localization-protected quantum order},\ }\href
  {https://doi.org/10.1103/PhysRevB.88.014206} {\bibfield  {journal} {\bibinfo
  {journal} {Phys. Rev. B}\ }\textbf {\bibinfo {volume} {88}},\ \bibinfo
  {pages} {014206} (\bibinfo {year} {2013})}\BibitemShut {NoStop}%
\bibitem [{\citenamefont {Laflorencie}\ \emph {et~al.}(2022)\citenamefont
  {Laflorencie}, \citenamefont {Lemari\'e},\ and\ \citenamefont
  {Mac\'e}}]{Laflorencie_2020}%
  \BibitemOpen
  \bibfield  {author} {\bibinfo {author} {\bibfnamefont {N.}~\bibnamefont
  {Laflorencie}}, \bibinfo {author} {\bibfnamefont {G.}~\bibnamefont
  {Lemari\'e}},\ and\ \bibinfo {author} {\bibfnamefont {N.}~\bibnamefont
  {Mac\'e}},\ }\bibfield  {title} {\bibinfo {title} {Topological order in
  random interacting ising-majorana chains stabilized by many-body
  localization},\ }\href {https://doi.org/10.1103/PhysRevResearch.4.L032016}
  {\bibfield  {journal} {\bibinfo  {journal} {Phys. Rev. Res.}\ }\textbf
  {\bibinfo {volume} {4}},\ \bibinfo {pages} {L032016} (\bibinfo {year}
  {2022})}\BibitemShut {NoStop}%
\bibitem [{\citenamefont {Kells}\ \emph {et~al.}(2018)\citenamefont {Kells},
  \citenamefont {Moran},\ and\ \citenamefont {Meidan}}]{Kells2018}%
  \BibitemOpen
  \bibfield  {author} {\bibinfo {author} {\bibfnamefont {G.}~\bibnamefont
  {Kells}}, \bibinfo {author} {\bibfnamefont {N.}~\bibnamefont {Moran}},\ and\
  \bibinfo {author} {\bibfnamefont {D.}~\bibnamefont {Meidan}},\ }\bibfield
  {title} {\bibinfo {title} {Localization enhanced and degraded topological
  order in interacting $p$-wave wires},\ }\href
  {https://doi.org/10.1103/PhysRevB.97.085425} {\bibfield  {journal} {\bibinfo
  {journal} {Phys. Rev. B}\ }\textbf {\bibinfo {volume} {97}},\ \bibinfo
  {pages} {085425} (\bibinfo {year} {2018})}\BibitemShut {NoStop}%
\bibitem [{\citenamefont {Shapiro}\ and\ \citenamefont
  {Tauber}(2019)}]{shapiro2019strongly}%
  \BibitemOpen
  \bibfield  {author} {\bibinfo {author} {\bibfnamefont {J.}~\bibnamefont
  {Shapiro}}\ and\ \bibinfo {author} {\bibfnamefont {C.}~\bibnamefont
  {Tauber}},\ }\bibfield  {title} {\bibinfo {title} {Strongly disordered
  floquet topological systems},\ }in\ \href@noop {} {\emph {\bibinfo
  {booktitle} {Annales Henri Poincar{\'e}}}},\ Vol.~\bibinfo {volume} {20}\
  (\bibinfo {organization} {Springer},\ \bibinfo {year} {2019})\ pp.\ \bibinfo
  {pages} {1837--1875}\BibitemShut {NoStop}%
\bibitem [{\citenamefont {Shtanko}\ and\ \citenamefont
  {Movassagh}(2018)}]{Shtanko2018StabilityTopoDisorder}%
  \BibitemOpen
  \bibfield  {author} {\bibinfo {author} {\bibfnamefont {O.}~\bibnamefont
  {Shtanko}}\ and\ \bibinfo {author} {\bibfnamefont {R.}~\bibnamefont
  {Movassagh}},\ }\bibfield  {title} {\bibinfo {title} {Stability of
  periodically driven topological phases against disorder},\ }\href
  {https://doi.org/10.1103/PhysRevLett.121.126803} {\bibfield  {journal}
  {\bibinfo  {journal} {Phys. Rev. Lett.}\ }\textbf {\bibinfo {volume} {121}},\
  \bibinfo {pages} {126803} (\bibinfo {year} {2018})}\BibitemShut {NoStop}%
\bibitem [{\citenamefont {Farrell}\ and\ \citenamefont
  {Pereg-Barnea}(2016{\natexlab{b}})}]{Farrell2018EdgeStateDisorder}%
  \BibitemOpen
  \bibfield  {author} {\bibinfo {author} {\bibfnamefont {A.}~\bibnamefont
  {Farrell}}\ and\ \bibinfo {author} {\bibfnamefont {T.}~\bibnamefont
  {Pereg-Barnea}},\ }\bibfield  {title} {\bibinfo {title} {Edge-state transport
  in floquet topological insulators},\ }\href
  {https://doi.org/10.1103/PhysRevB.93.045121} {\bibfield  {journal} {\bibinfo
  {journal} {Phys. Rev. B}\ }\textbf {\bibinfo {volume} {93}},\ \bibinfo
  {pages} {045121} (\bibinfo {year} {2016}{\natexlab{b}})}\BibitemShut
  {NoStop}%
\bibitem [{\citenamefont {Titum}\ \emph {et~al.}(2015)\citenamefont {Titum},
  \citenamefont {Lindner}, \citenamefont {Rechtsman},\ and\ \citenamefont
  {Refael}}]{Titum2015DisorderInduceTopology}%
  \BibitemOpen
  \bibfield  {author} {\bibinfo {author} {\bibfnamefont {P.}~\bibnamefont
  {Titum}}, \bibinfo {author} {\bibfnamefont {N.~H.}\ \bibnamefont {Lindner}},
  \bibinfo {author} {\bibfnamefont {M.~C.}\ \bibnamefont {Rechtsman}},\ and\
  \bibinfo {author} {\bibfnamefont {G.}~\bibnamefont {Refael}},\ }\bibfield
  {title} {\bibinfo {title} {Disorder-induced floquet topological insulators},\
  }\href {https://doi.org/10.1103/PhysRevLett.114.056801} {\bibfield  {journal}
  {\bibinfo  {journal} {Phys. Rev. Lett.}\ }\textbf {\bibinfo {volume} {114}},\
  \bibinfo {pages} {056801} (\bibinfo {year} {2015})}\BibitemShut {NoStop}%
\bibitem [{\citenamefont {Decker}\ \emph {et~al.}(2020)\citenamefont {Decker},
  \citenamefont {Karrasch}, \citenamefont {Eisert},\ and\ \citenamefont
  {Kennes}}]{PhysRevLett.124.190601}%
  \BibitemOpen
  \bibfield  {author} {\bibinfo {author} {\bibfnamefont {K.~S.~C.}\
  \bibnamefont {Decker}}, \bibinfo {author} {\bibfnamefont {C.}~\bibnamefont
  {Karrasch}}, \bibinfo {author} {\bibfnamefont {J.}~\bibnamefont {Eisert}},\
  and\ \bibinfo {author} {\bibfnamefont {D.~M.}\ \bibnamefont {Kennes}},\
  }\bibfield  {title} {\bibinfo {title} {Floquet engineering topological
  many-body localized systems},\ }\href
  {https://doi.org/10.1103/PhysRevLett.124.190601} {\bibfield  {journal}
  {\bibinfo  {journal} {Phys. Rev. Lett.}\ }\textbf {\bibinfo {volume} {124}},\
  \bibinfo {pages} {190601} (\bibinfo {year} {2020})}\BibitemShut {NoStop}%
\bibitem [{\citenamefont {Arnoldi}(1951)}]{Arnoldi}%
  \BibitemOpen
  \bibfield  {author} {\bibinfo {author} {\bibfnamefont {W.~E.}\ \bibnamefont
  {Arnoldi}},\ }\bibfield  {title} {\bibinfo {title} {The principle of
  minimized iterations in the solution of the matrix eigenvalue problem},\
  }\href@noop {} {\bibfield  {journal} {\bibinfo  {journal} {Quarterly of
  Applied Mathematics}\ }\textbf {\bibinfo {volume} {9}},\ \bibinfo {pages}
  {17} (\bibinfo {year} {1951})}\BibitemShut {NoStop}%
\bibitem [{\citenamefont {Yates}\ \emph {et~al.}(2019)\citenamefont {Yates},
  \citenamefont {Essler},\ and\ \citenamefont {Mitra}}]{Yates_2019}%
  \BibitemOpen
  \bibfield  {author} {\bibinfo {author} {\bibfnamefont {D.~J.}\ \bibnamefont
  {Yates}}, \bibinfo {author} {\bibfnamefont {F.~H.~L.}\ \bibnamefont
  {Essler}},\ and\ \bibinfo {author} {\bibfnamefont {A.}~\bibnamefont
  {Mitra}},\ }\bibfield  {title} {\bibinfo {title} {Almost strong
  ($0,\ensuremath{\pi}$) edge modes in clean interacting one-dimensional
  floquet systems},\ }\href {https://doi.org/10.1103/PhysRevB.99.205419}
  {\bibfield  {journal} {\bibinfo  {journal} {Phys. Rev. B}\ }\textbf {\bibinfo
  {volume} {99}},\ \bibinfo {pages} {205419} (\bibinfo {year}
  {2019})}\BibitemShut {NoStop}%
\bibitem [{\citenamefont {Yates}\ \emph
  {et~al.}(2020{\natexlab{a}})\citenamefont {Yates}, \citenamefont {Abanov},\
  and\ \citenamefont {Mitra}}]{Yates_PRL_2020}%
  \BibitemOpen
  \bibfield  {author} {\bibinfo {author} {\bibfnamefont {D.~J.}\ \bibnamefont
  {Yates}}, \bibinfo {author} {\bibfnamefont {A.~G.}\ \bibnamefont {Abanov}},\
  and\ \bibinfo {author} {\bibfnamefont {A.}~\bibnamefont {Mitra}},\ }\bibfield
   {title} {\bibinfo {title} {Lifetime of almost strong edge-mode operators in
  one-dimensional, interacting, symmetry protected topological phases},\ }\href
  {https://doi.org/10.1103/PhysRevLett.124.206803} {\bibfield  {journal}
  {\bibinfo  {journal} {Phys. Rev. Lett.}\ }\textbf {\bibinfo {volume} {124}},\
  \bibinfo {pages} {206803} (\bibinfo {year} {2020}{\natexlab{a}})}\BibitemShut
  {NoStop}%
\bibitem [{\citenamefont {Yates}\ \emph
  {et~al.}(2020{\natexlab{b}})\citenamefont {Yates}, \citenamefont {Abanov},\
  and\ \citenamefont {Mitra}}]{Yates_2020}%
  \BibitemOpen
  \bibfield  {author} {\bibinfo {author} {\bibfnamefont {D.~J.}\ \bibnamefont
  {Yates}}, \bibinfo {author} {\bibfnamefont {A.~G.}\ \bibnamefont {Abanov}},\
  and\ \bibinfo {author} {\bibfnamefont {A.}~\bibnamefont {Mitra}},\ }\bibfield
   {title} {\bibinfo {title} {Dynamics of almost strong edge modes in spin
  chains away from integrability},\ }\href
  {https://doi.org/10.1103/PhysRevB.102.195419} {\bibfield  {journal} {\bibinfo
   {journal} {Phys. Rev. B}\ }\textbf {\bibinfo {volume} {102}},\ \bibinfo
  {pages} {195419} (\bibinfo {year} {2020}{\natexlab{b}})}\BibitemShut
  {NoStop}%
\bibitem [{\citenamefont {Yates}\ and\ \citenamefont
  {Mitra}(2021)}]{Yates_2021}%
  \BibitemOpen
  \bibfield  {author} {\bibinfo {author} {\bibfnamefont {D.~J.}\ \bibnamefont
  {Yates}}\ and\ \bibinfo {author} {\bibfnamefont {A.}~\bibnamefont {Mitra}},\
  }\bibfield  {title} {\bibinfo {title} {Strong and almost strong modes of
  floquet spin chains in krylov subspaces},\ }\href
  {https://doi.org/10.1103/PhysRevB.104.195121} {\bibfield  {journal} {\bibinfo
   {journal} {Phys. Rev. B}\ }\textbf {\bibinfo {volume} {104}},\ \bibinfo
  {pages} {195121} (\bibinfo {year} {2021})}\BibitemShut {NoStop}%
\bibitem [{\citenamefont {Yates}\ \emph {et~al.}(2022)\citenamefont {Yates},
  \citenamefont {Abanov},\ and\ \citenamefont {Mitra}}]{Yates_2022}%
  \BibitemOpen
  \bibfield  {author} {\bibinfo {author} {\bibfnamefont {D.}~\bibnamefont
  {Yates}}, \bibinfo {author} {\bibfnamefont {A.}~\bibnamefont {Abanov}},\ and\
  \bibinfo {author} {\bibfnamefont {A.}~\bibnamefont {Mitra}},\ }\bibfield
  {title} {\bibinfo {title} {Long-lived period-doubled edge modes of
  interacting and disorder-free floquet spin chains},\ }\href
  {https://doi.org/10.1038/s42005-022-00818-1} {\bibfield  {journal} {\bibinfo
  {journal} {Commun Phys}\ }\textbf {\bibinfo {volume} {5}},\ \bibinfo {pages}
  {43} (\bibinfo {year} {2022})}\BibitemShut {NoStop}%
\bibitem [{\citenamefont {Yeh}\ \emph {et~al.}()\citenamefont {Yeh},
  \citenamefont {Rosch},\ and\ \citenamefont {Mitra}}]{Yeh_2023}%
  \BibitemOpen
  \bibfield  {author} {\bibinfo {author} {\bibfnamefont {H.-C.}\ \bibnamefont
  {Yeh}}, \bibinfo {author} {\bibfnamefont {A.}~\bibnamefont {Rosch}},\ and\
  \bibinfo {author} {\bibfnamefont {A.}~\bibnamefont {Mitra}},\ }\bibfield
  {title} {\bibinfo {title} {Decay rates of almost strong modes in floquet spin
  chains beyond fermi's golden rule},\ }\bibfield  {journal} {\bibinfo
  {journal} {arXiv/2305.04980}\ }\href
  {https://doi.org/10.1103/PhysRevB.88.014206}
  {10.1103/PhysRevB.88.014206}\BibitemShut {NoStop}%
\bibitem [{\citenamefont {Pientka}\ \emph {et~al.}(2013)\citenamefont
  {Pientka}, \citenamefont {Romito}, \citenamefont {Duckheim}, \citenamefont
  {Oreg},\ and\ \citenamefont {von Oppen}}]{Pientka_2013}%
  \BibitemOpen
  \bibfield  {author} {\bibinfo {author} {\bibfnamefont {F.}~\bibnamefont
  {Pientka}}, \bibinfo {author} {\bibfnamefont {A.}~\bibnamefont {Romito}},
  \bibinfo {author} {\bibfnamefont {M.}~\bibnamefont {Duckheim}}, \bibinfo
  {author} {\bibfnamefont {Y.}~\bibnamefont {Oreg}},\ and\ \bibinfo {author}
  {\bibfnamefont {F.}~\bibnamefont {von Oppen}},\ }\bibfield  {title} {\bibinfo
  {title} {Signatures of topological phase transitions in mesoscopic
  superconducting rings},\ }\href
  {https://doi.org/10.1088/1367-2630/15/2/025001} {\bibfield  {journal}
  {\bibinfo  {journal} {New Journal of Physics}\ }\textbf {\bibinfo {volume}
  {15}},\ \bibinfo {pages} {025001} (\bibinfo {year} {2013})}\BibitemShut
  {NoStop}%
\bibitem [{\citenamefont {Gergs}\ \emph {et~al.}(2016)\citenamefont {Gergs},
  \citenamefont {Fritz},\ and\ \citenamefont {Schuricht}}]{Gergs_2016}%
  \BibitemOpen
  \bibfield  {author} {\bibinfo {author} {\bibfnamefont {N.~M.}\ \bibnamefont
  {Gergs}}, \bibinfo {author} {\bibfnamefont {L.}~\bibnamefont {Fritz}},\ and\
  \bibinfo {author} {\bibfnamefont {D.}~\bibnamefont {Schuricht}},\ }\bibfield
  {title} {\bibinfo {title} {Topological order in the kitaev/majorana chain in
  the presence of disorder and interactions},\ }\href
  {https://doi.org/10.1103/PhysRevB.93.075129} {\bibfield  {journal} {\bibinfo
  {journal} {Phys. Rev. B}\ }\textbf {\bibinfo {volume} {93}},\ \bibinfo
  {pages} {075129} (\bibinfo {year} {2016})}\BibitemShut {NoStop}%
\bibitem [{\citenamefont {Hegde}\ and\ \citenamefont
  {Vishveshwara}(2016)}]{Hegde_2016}%
  \BibitemOpen
  \bibfield  {author} {\bibinfo {author} {\bibfnamefont {S.~S.}\ \bibnamefont
  {Hegde}}\ and\ \bibinfo {author} {\bibfnamefont {S.}~\bibnamefont
  {Vishveshwara}},\ }\bibfield  {title} {\bibinfo {title} {Majorana
  wave-function oscillations, fermion parity switches, and disorder in kitaev
  chains},\ }\href {https://doi.org/10.1103/PhysRevB.94.115166} {\bibfield
  {journal} {\bibinfo  {journal} {Phys. Rev. B}\ }\textbf {\bibinfo {volume}
  {94}},\ \bibinfo {pages} {115166} (\bibinfo {year} {2016})}\BibitemShut
  {NoStop}%
\bibitem [{\citenamefont {Pan}\ and\ \citenamefont
  {Das~Sarma}(2021{\natexlab{b}})}]{Pan_2021}%
  \BibitemOpen
  \bibfield  {author} {\bibinfo {author} {\bibfnamefont {H.}~\bibnamefont
  {Pan}}\ and\ \bibinfo {author} {\bibfnamefont {S.}~\bibnamefont
  {Das~Sarma}},\ }\bibfield  {title} {\bibinfo {title} {Disorder effects on
  majorana zero modes: Kitaev chain versus semiconductor nanowire},\ }\href
  {https://doi.org/10.1103/PhysRevB.103.224505} {\bibfield  {journal} {\bibinfo
   {journal} {Phys. Rev. B}\ }\textbf {\bibinfo {volume} {103}},\ \bibinfo
  {pages} {224505} (\bibinfo {year} {2021}{\natexlab{b}})}\BibitemShut
  {NoStop}%
\bibitem [{\citenamefont {Vishwanath}\ and\ \citenamefont
  {M\"{u}ller}(2008)}]{Recbook}%
  \BibitemOpen
  \bibfield  {author} {\bibinfo {author} {\bibfnamefont {V.}~\bibnamefont
  {Vishwanath}}\ and\ \bibinfo {author} {\bibfnamefont {G.}~\bibnamefont
  {M\"{u}ller}},\ }\href@noop {} {\bibfield  {journal} {\bibinfo  {journal}
  {{\sl The Recursion Method: Applications to Many-Body Dynamics}, Springer,
  New York}\ } (\bibinfo {year} {2008})}\BibitemShut {NoStop}%
\bibitem [{\citenamefont {Horn}\ and\ \citenamefont
  {Johnson}(2013)}]{Horn_2013}%
  \BibitemOpen
  \bibfield  {author} {\bibinfo {author} {\bibfnamefont {R.~A.}\ \bibnamefont
  {Horn}}\ and\ \bibinfo {author} {\bibfnamefont {C.~R.}\ \bibnamefont
  {Johnson}},\ }\href@noop {} {\emph {\bibinfo {title} {Matrix Analysis}}}\
  (\bibinfo  {publisher} {Cambridge University Press},\ \bibinfo {address}
  {Cambridge, UK},\ \bibinfo {year} {2013})\BibitemShut {NoStop}%
\bibitem [{\citenamefont {Su}\ \emph {et~al.}(1979)\citenamefont {Su},
  \citenamefont {Schrieffer},\ and\ \citenamefont {Heeger}}]{SSH}%
  \BibitemOpen
  \bibfield  {author} {\bibinfo {author} {\bibfnamefont {W.~P.}\ \bibnamefont
  {Su}}, \bibinfo {author} {\bibfnamefont {J.~R.}\ \bibnamefont {Schrieffer}},\
  and\ \bibinfo {author} {\bibfnamefont {A.~J.}\ \bibnamefont {Heeger}},\
  }\bibfield  {title} {\bibinfo {title} {Solitons in polyacetylene},\ }\href
  {https://doi.org/10.1103/PhysRevLett.42.1698} {\bibfield  {journal} {\bibinfo
   {journal} {Phys. Rev. Lett.}\ }\textbf {\bibinfo {volume} {42}},\ \bibinfo
  {pages} {1698} (\bibinfo {year} {1979})}\BibitemShut {NoStop}%
\bibitem [{\citenamefont {Casini}\ and\ \citenamefont
  {Huerta}(2004)}]{Casini_2004}%
  \BibitemOpen
  \bibfield  {author} {\bibinfo {author} {\bibfnamefont {H.}~\bibnamefont
  {Casini}}\ and\ \bibinfo {author} {\bibfnamefont {M.}~\bibnamefont
  {Huerta}},\ }\bibfield  {title} {\bibinfo {title} {A finite entanglement
  entropy and the c-theorem},\ }\href
  {https://doi.org/https://doi.org/10.1016/j.physletb.2004.08.072} {\bibfield
  {journal} {\bibinfo  {journal} {Physics Letters B}\ }\textbf {\bibinfo
  {volume} {600}},\ \bibinfo {pages} {142} (\bibinfo {year}
  {2004})}\BibitemShut {NoStop}%
\bibitem [{\citenamefont {Levy}\ and\ \citenamefont
  {Goldstein}(2019)}]{Levy_2019}%
  \BibitemOpen
  \bibfield  {author} {\bibinfo {author} {\bibfnamefont {L.}~\bibnamefont
  {Levy}}\ and\ \bibinfo {author} {\bibfnamefont {M.}~\bibnamefont
  {Goldstein}},\ }\bibfield  {title} {\bibinfo {title} {Entanglement and
  disordered-enhanced topological phase in the kitaev chain},\ }\bibfield
  {journal} {\bibinfo  {journal} {Universe}\ }\textbf {\bibinfo {volume} {5}},\
  \href {https://doi.org/10.3390/universe5010033} {10.3390/universe5010033}
  (\bibinfo {year} {2019})\BibitemShut {NoStop}%
\bibitem [{\citenamefont {Li}\ \emph {et~al.}(2009)\citenamefont {Li},
  \citenamefont {Chu}, \citenamefont {Jain},\ and\ \citenamefont
  {Shen}}]{Li_2009}%
  \BibitemOpen
  \bibfield  {author} {\bibinfo {author} {\bibfnamefont {J.}~\bibnamefont
  {Li}}, \bibinfo {author} {\bibfnamefont {R.-L.}\ \bibnamefont {Chu}},
  \bibinfo {author} {\bibfnamefont {J.~K.}\ \bibnamefont {Jain}},\ and\
  \bibinfo {author} {\bibfnamefont {S.-Q.}\ \bibnamefont {Shen}},\ }\bibfield
  {title} {\bibinfo {title} {Topological anderson insulator},\ }\href
  {https://doi.org/10.1103/PhysRevLett.102.136806} {\bibfield  {journal}
  {\bibinfo  {journal} {Phys. Rev. Lett.}\ }\textbf {\bibinfo {volume} {102}},\
  \bibinfo {pages} {136806} (\bibinfo {year} {2009})}\BibitemShut {NoStop}%
\end{thebibliography}%
						
					\end{document}